\documentclass[iop,numberedappendix,twocolappendix,twocolumn]{emulateapj}

\usepackage{amsmath,amssymb,amstext}
\usepackage{float} 
\usepackage{url}
\usepackage{natbib}
\usepackage{graphicx}
\usepackage{color}
\usepackage{rotating}
\usepackage{lineno}
\usepackage[export]{adjustbox}
\usepackage{comment}

\usepackage[breaklinks,colorlinks,citecolor=blue,linkcolor=magenta]{hyperref} 

\usepackage[all]{hypcap} 

\newcommand{\Planck}{\textit{Planck }}

\newcommand{\FSN}{\mathrm{FSN}}

\providecommand{\sorthelp}[1]{}

\shorttitle{Translucent Molecular Cloud Polarization Spectrum}
\shortauthors{Ashton et al.}

\begin{document}

\title{First Observation of the Submillimeter Polarization Spectrum in a Translucent Molecular Cloud}
\author{
Peter C. Ashton\altaffilmark{1},
Peter A. R. Ade\altaffilmark{2},
Francesco E. Angil\`e\altaffilmark{3},
Steven J. Benton\altaffilmark{4},
Mark J. Devlin\altaffilmark{3},
Bradley Dober\altaffilmark{5},
Laura M. Fissel\altaffilmark{6},
Yasuo Fukui\altaffilmark{7},
Nicholas Galitzki\altaffilmark{8},
Natalie N. Gandilo\altaffilmark{9},
Jeffrey Klein\altaffilmark{3},
Andrei L. Korotkov\altaffilmark{10},
Zhi-Yun Li\altaffilmark{11},
Peter G. Martin\altaffilmark{12},
Tristan G. Matthews\altaffilmark{1},
Lorenzo Moncelsi\altaffilmark{13},
Fumitaka Nakamura\altaffilmark{14},
Calvin B. Netterfield\altaffilmark{15,16},
Giles Novak\altaffilmark{1},
Enzo Pascale\altaffilmark{2},
Fr{\'e}d{\'e}rick Poidevin\altaffilmark{17,18},
Fabio P. Santos\altaffilmark{1},
Giorgio Savini\altaffilmark{19},
Douglas Scott\altaffilmark{20},
Jamil A. Shariff\altaffilmark{12},
Juan D. Soler\altaffilmark{21},
Nicholas E. Thomas\altaffilmark{22},
Carole E. Tucker\altaffilmark{2},
Gregory S. Tucker\altaffilmark{11},
Derek Ward-Thompson\altaffilmark{23}
}

\altaffiltext{1}{Center for Interdisciplinary Exploration and Research in Astrophysics (CIERA) and Department\ of Physics \& Astronomy, Northwestern University, 2145 Sheridan Road, Evanston} 
\altaffiltext{2}{Cardiff University, School of Physics \& Astronomy, Queens Buildings, The Parade, Cardiff, CF24 3AA, U.K.}
\altaffiltext{3}{Department of Physics \& Astronomy, University of Pennsylvania, 209 South 33rd Street, Philadelphia, PA, 19104, U.S.A.}
\altaffiltext{4}{Department of Physics, Princeton University, Jadwin Hall, Princeton, NJ 08544, U.S.A.}
\altaffiltext{5}{National Institute of Standards and Technology, 325 Broadway, Boulder, CO 80305, U.S.A.}
\altaffiltext{6}{National Radio Astronomy Observatory, 520 Edgemont Rd. Charlottesville, VA 22903, U.S.A.}
\altaffiltext{7}{Department of Physics and Astrophysics, Nagoya University, Nagoya 464-8602, Japan}
\altaffiltext{8}{Center for Astrophysics and Space Sciences, University of California San Diego, M/C 0424, 9500 Gilman Drive, La Jolla, CA 92093, U.S.A.}
\altaffiltext{9}{Department of Physics and Astronomy, Johns Hopkins University, 3701 San Martin Drive, Baltimore, Maryland, USA}
\altaffiltext{10}{Department of Physics, Brown University, 182 Hope Street, Providence, RI, 02912, U.S.A.}
\altaffiltext{11}{Department of Astronomy, University of Virginia, 530 McCormick Rd, Charlottesville, VA 22904, U.S.A.}
\altaffiltext{12}{CITA, University of Toronto, 60 St. George St., Toronto, ON M5S 3H8, Canada}
\altaffiltext{13}{California Institute of Technology, 1200 E. California Blvd., Pasadena, CA, 91125, U.S.A.}
\altaffiltext{14}{National Astronomical Observatory, Mitaka, Tokyo 181-8588, Japan}
\altaffiltext{15}{Department of Astronomy \& Astrophysics, University of Toronto, 50 St. George Street Toronto, ON M5S 3H4, Canada}
\altaffiltext{16}{Department of Physics, University of Toronto, 60 St. George Street Toronto, ON M5S 1A7, Canada}
\altaffiltext{17}{Instituto de Astrof\'{i}sica de Canarias (IAC), E-38205 La Laguna, Tenerife, Spain}
\altaffiltext{18}{Universidad de La Laguna, Depto. Astrof\'{i}sica, E-38206 La Laguna, Tenerife, Spain}
\altaffiltext{19}{Department of Physics \& Astronomy, University College London, Gower Street, London, WC1E 6BT, U.K.}
\altaffiltext{20}{Department of Physics \& Astronomy, University of British Columbia, 6224 Agricultural Road, Vancouver, BC V6T 1Z1, Canada}
\altaffiltext{21}{Max-Planck-Institute for Astronomy, K\"{o}nigstuhl 17, 69117, Heidelberg, Germany}
\altaffiltext{22}{NASA/Goddard Space Flight Center, Greenbelt, MD 20771, U.S.A.}
\altaffiltext{23}{Jeremiah Horrocks Institute, University of Central Lancashire, PR1 2HE, U.K.}

\begin{abstract} 
Polarized emission from aligned dust is a crucial tool for studies of magnetism in the ISM and a troublesome contaminant for studies of CMB polarization. In each case, an understanding of the significance of the polarization signal requires well-calibrated physical models of dust grains. Despite decades of progress in theory and observation, polarized dust models remain largely underconstrained. During its 2012 flight, the balloon-borne telescope BLASTPol obtained simultaneous broad-band polarimetric maps of a translucent molecular cloud at 250, 350, and 500 $\mu$m. Combining these data with polarimetry from the \Planck 850 $\mu$m band, we have produced a submillimeter polarization spectrum for a cloud of this type for the first time. We find the polarization degree to be largely constant across the four bands. This result introduces a new observable with the potential to place strong empirical constraints on ISM dust polarization models in a previously inaccessible density regime. Comparing with models by Draine and Fraisse (2009), our result disfavors two of their models for which all polarization arises due only to aligned silicate grains. By creating simple models for polarized emission in a translucent cloud, we verify that extinction within the cloud should have only a small effect on the polarization spectrum shape compared to the diffuse ISM. Thus we expect the measured polarization spectrum to be a valid check on diffuse ISM dust models. The general flatness of the observed polarization spectrum suggests a challenge to models where temperature and alignment degree are strongly correlated across major dust components.
\end{abstract}


\maketitle
\section{Introduction}
For nearly 90 years, interstellar dust has been known to pervade the volume of the interstellar medium (ISM; \citealt{Trumpler30}). Though the presence of dust has been well-established, many of the physical properties of dust populations remain poorly determined. To this end, future dust grain models will be better constrained by comparisons to observations of dust polarized thermal emission than by comparisons to total power alone. This polarized signal, arising due to the tendency of spinning dust grains to align with their long axes perpendicular to the local magnetic field, will vary as a function of grain composition, size, shape, and temperature, among other parameters (see reviews by \citealt{Lazarian07} and \citealt{Andersson15}). Thus new observations of dust polarized emission, especially over a large range of wavelengths and across a range of ISM environments, will help provide better empirical constraints on dust properties. 

In particular, the polarization spectrum, the variation of polarization degree with wavelength (i.e.\ $p(\lambda)$), carries information about the optical properties of grain material, the efficiency of alignment processes on various populations, and the interstellar radiation field (ISRF) to which the dust is exposed. The availability of more precise dust models for all environments will aid in answering important questions in other areas of astrophysics. For example, to understand the dust polarization signal as a tracer of magnetism in the ISM or to isolate \textit{B}-mode polarization in the cosmic microwave background (CMB) from Galactic dust foregrounds, constrained models of polarized dust emission are necessary to draw conclusions from observations. 

Translucent molecular clouds were first proposed as a classification of interstellar molecular gas by \citet{vanDishoeck89}. Originally used to describe molecular clouds with 2\,mag $ < A_V^{\mathrm{tot}} < $ 10\,mag, \citet{Snow06} propose a definition based on the state of carbon within the cloud ($<$ 50\% of carbon atoms present as C$^+$ and $<$ 90\% of carbon atoms in CO molecules). Intuitively, translucent molecular clouds are intermediate density structures between the diffuse ISM and dense molecular clouds. In these regions, the ISRF plays a role in determining the cloud thermal structure and chemistry, although it becomes increasingly more attenuated deeper into the cloud. For the present work, this type of cloud represents a new column density regime in which to study the submillimeter polarization spectrum.

In this paper, we present the first determination of the submillimeter dust polarization spectrum in a translucent molecular cloud. However, the observed polarization degree may be uniformly reduced across the wavelength range (e.g.\ due to the inclination angle of the magnetic field to the line of sight; see \citealt{Andersson15} for a review). Thus we will only discuss the polarization spectrum normalized to its value at a single wavelength: $p(\lambda)/p(\lambda_0)$. This normalization effectively removes the dependence on unknown factors, and we will interpret spectrum shapes rather than absolute polarization degree.

Early measurements of the submillimeter polarization spectrum  by \citet{Hildebrand99}, \citet{Vaillancourt08}, \citet{VM12}, and \citet{Zeng13} were limited to relatively warm and bright star-forming regions with embedded high-mass young stellar objects (YSOs). By aggregating results for several targets, these authors found a polarization spectrum with a pronounced minimum near 350\,$\mu$m, with the polarization degree a factor of a few higher in bands shortward of $\sim$100\,$\mu$m and longward of $\sim$1\,mm. \citet{Gandilo16} first measured the polarization spectrum in a dense molecular cloud without embedded high-mass YSOs and found it to be largely flat between 250\,$\mu$m and 850\,$\mu$m. For more diffuse lines of sight, the \Planck observatory investigated the polarization spectrum for wavelengths from 850\,$\mu$m to 4.3\,mm, and found it to be flat or decreasing with increasing wavelength \citep{planck2014-XXII}. In this work, we will extend submillimeter polarization spectrum studies to translucent molecular clouds for the first time.   

The data we use were obtained using the Balloon-borne Large Aperture Submillimeter Telescope for Polarimetry (BLASTPol) during its 2012 flight aboard a high-altitude balloon from McMurdo Station, Antarctica \citep{Galitzki14_BPOL}. Among the targets it observed was a region in the Vela Molecular Ridge, for which BLASTPol produced maps of Stokes \textit{I}, \textit{Q}, and \textit{U} at 250\,$\mu$m, 350\,$\mu$m, and 500\,$\mu$m. This data set has been used in four previous studies (\citealt{Fissel16}; \citealt{Gandilo16}; \citealt{Santos17}; \citealt{Soler17}, submitted), all of which focused on various aspects of the polarization signal in the Vela C molecular cloud. In the present work we instead focus on lower column density lines of sight in the region away from Vela C.

Additionally, we use \Planck HFI \citep{planck2014-a09} polarimetry maps at 850\,$\mu$m or 353\,GHz \citep{planck2014-XIX}, obtained via the Planck Legacy Archive portal.\footnote[1]{\url{http://pla.esac.esa.int/pla/}} We also use the products of the \Planck thermal dust model \citep{planck2013-p06b}, which produces all sky maps of dust optical depth ($\tau_{353}$), temperature (\textit{T}), and spectral index ($\beta$) from modified blackbody fits. The \Planck maps of the Vela Molecular Ridge region are projected and resampled onto a Cartesian grid using the gnomonic projection procedure described in \citet{Paradis12}. The present analysis is performed on these projected maps. The selected region is small enough and is located at sufficiently low Galactic latitudes that this projection does not significantly impact our study.

The paper is organized as follows. In Section 2 we describe the data products and analysis procedures used for isolating the cloud polarization signal. In Section 3 we present the polarization spectrum and discuss its associated statistical and systematic errors. In Section 4 we compare our results to models of the polarization spectrum from the literature. In Section 5 we generate simple analytic models of polarization spectra in order to validate our understanding of the physical effects at play. In Section 6 we discuss and summarize our findings. Appendices A and B provide details on our error analyses and on our modeling methodology, respectively.

\newpage

\section{Filtered BLASTPol Maps and Target Selection}

\subsection{Description of BLASTPol systematic errors}
In order to understand the challenge of extracting polarization spectrum information from the BLASTPol maps, it is first necessary to consider the method by which the instrument makes its measurements. As the BLASTPol telescope scans across the sky, it makes differential measurements of power. It is therefore subject to low-frequency time-variable drifts in amplifier gain, optics temperature, and detector noise properties, broadly referred to as ``1/\textit{f} noise''. The results of these effects show up as variations in the final Stokes parameter maps. The \texttt{TOAST} mapmaking software (\citealt{Cantalupo10}; \citealt{Fissel16}) acts to minimize the influence of 1/\textit{f} noise in map regions with good cross-linking (areas where scans cross the region at many different parallactic angles). However the Vela Molecular Ridge map has ``wings" where there is little cross-linking and therefore systematic errors are dominant. The effect of these systematic errors is visible in the top right panel of Figure \ref{fig:bp_filt_resid}, a residual map which shows the difference between BLASTPol Stokes $Q_{250}$ and a scaled version of \Planck $Q_{850}$ (See Equation (\ref{eq2})). These regions, near the corners of the scanned area, stand in contrast with the region used in the above-mentioned studies on Vela C, which have superior cross-linking and signal to noise ratio (S/N). For this reason we must first pursue the removal of the spurious systematic 1/\textit{f} noise before advancing to analysis and interpretation of the polarization maps.

\subsection{Spatial filtering method} \label{filt_method}
Because the contribution to the Stokes parameter maps from 1/\textit{f} noise appears as a large-scale variation across the map area, a high-pass spatial filter can effectively remove systematic errors while preserving information on the scale of physical structures of interest. We therefore apply a high-pass filter to the Fourier transform of our Stokes parameter maps, specifically employing a 2D Butterworth filter:
\begin{equation}\label{eq1}
G(k) = 1 - \frac{1}{\sqrt{1 + (\frac{k}{k_0})^{2n}}}
\end{equation}
where k is a scaled 2D spatial frequency. The filter is characterized by two parameters: the cutoff mode ($k_0$) sets the scale of the filter's half-power point, and the order (\textit{n}) sets the ``sharpness" of the filter roll-off. In the limit $n \to \infty$, $G(k)$ approaches a step function with a stopband from \textit{k} = 0 to $k_0$. 

\begin{figure*}[ht]
\includegraphics[trim={0.5in, 0.25in, 0.5in, 0}, clip, width=0.8\textwidth, center]{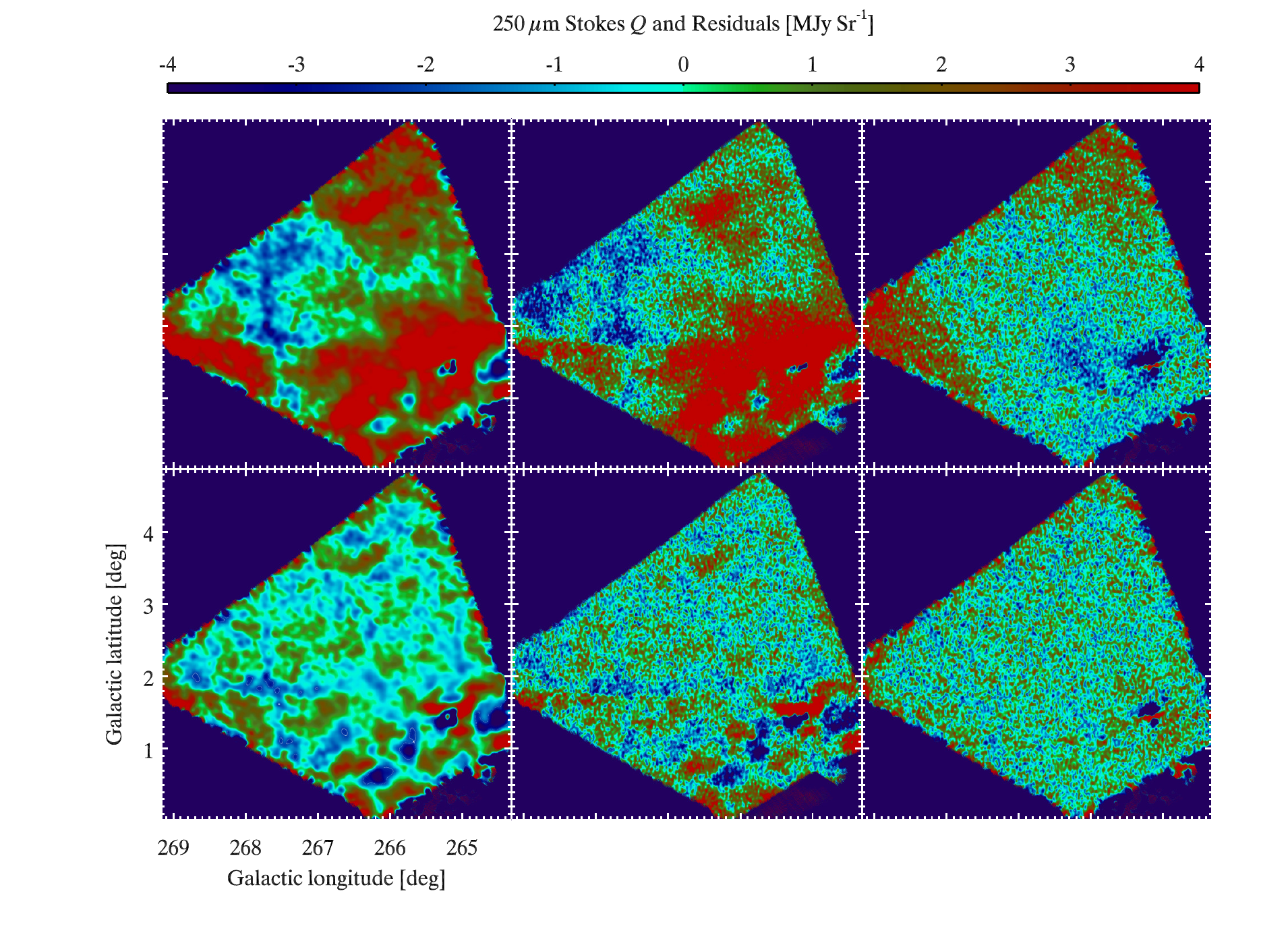}
\caption{From top left to bottom right: BLASTPol $Q_{250}$, corresponding flat-spectrum normalized $Q_{250}^{\FSN}$, $Q_{250} - Q_{250}^{\FSN}$ residual, high-pass filtered $Q_{250}$, high-pass filtered $Q_{250}^{\FSN}$, and residual of high-pass filtered maps.\label{fig:bp_filt_resid}}
\end{figure*}

Anticipating a comparison between BLASTPol and \Planck 850\,$\mu$m (353\,GHz) polarimetry maps, prior to applying the spatial filter, we smooth the BLASTPol maps to the \Planck resolution. The BLASTPol Stokes \textit{Q} and \textit{U} maps in each of the three wavelength bands are convolved with an appropriate 2D Gaussian kernel such that the resulting maps have a resolution of $4.8^{\prime}$. We then pad the area surrounding the region mapped by BLASTPol with zero values and trim the maps to a square region 4.86$^{\circ}$ on a side. This square is the box outlining each panel in Figure \ref{fig:bp_filt_resid}. We then filter this set of ``raw" maps in Fourier space using the Butterworth high-pass filter in Equation (\ref{eq1}). Initially we set $k_0$ = 4 in order to exclude the largest scale 1/\textit{f} wave that is visible by eye, and we set \textit{n} = 6 in order to sharply separate the low-order modes from the high-order modes without introducing high frequency ringing artifacts due to the filtering. The angular wavelength associated with a given scaled spatial frequency $k$ depends on the size of the square region in the spatial domain, and in our case is given by
\begin{equation} \label{eq_angscale}
	\theta(k) = \frac{4.86^{\circ}}{k}.
\end{equation}
Thus a choice of $k_0$ = 4 is equivalent to removing Fourier components with wavelengths larger than 1.215$^{\circ}$. A more complete justification of the choice of parameters and an investigation into the sensitivity of our result to that choice will follow in Section \ref{derive_polspec_sec} and Appendix \ref{appendix_sys}.

\subsection{Comparison with \Planck polarimetry}
In order to compare with data from the \Planck 850\,$\mu$m polarimetry maps, we produce maps:
\begin{equation}\label{eq2}
Q^{\FSN}_\lambda = Q_{850}\frac{I^{\mathrm{mdl}}_\lambda}{I_{850}}
\end{equation}
\begin{equation}\label{eq3}
U^{\FSN}_\lambda = U_{850}\frac{I^{\mathrm{mdl}}_\lambda}{I_{850}}
\end{equation}
where $I_{850}$, $Q_{850}$, and $U_{850}$ are maps of the \Planck 850\,$\mu$m Stokes parameters, and $I^{\mathrm{mdl}}_\lambda$  is the map of total intensity described by the \Planck thermal dust model modified blackbody integrated over BLASTPol band $\lambda$. These ``flat spectrum normalized" (FSN) Stokes parameter maps can be understood as the polarization signal one would expect BLASTPol to observe in a given band under the assumption that the polarization degree and angle are the same in that band as in the \Planck 850\,$\mu$m band. The FSN maps will have none of scan-correlated systematic errors that affect the BLASTPol maps. Nevertheless, to facilitate comparison with BLASTPol, we then spatially filter the $Q^{\FSN}$ and $U^{\FSN}$ maps using the same high-pass filter applied to the BLASTPol maps.

Note that in the FSN constructions of Equations (\ref{eq2}) and (\ref{eq3}), we avoid reference to the BLASTPol Stokes $I$ maps, where we expect the 1/\textit{f} contamination to be most significant. In fact, the BLASTPol $I_\lambda$ maps have been calibrated (gain and zero-point) by comparing with $I^{\mathrm{mdl}}_\lambda$ in regions with good cross-linking, so we expect $I^{\mathrm{mdl}}_\lambda$ to be a valid estimation of $I_\lambda$, without 1/\textit{f} contamination. 

Examination of pairs of BLASTPol-observed and \textit{Planck}-modeled FSN Stokes parameter maps reveals similar polarization structures. As an example, Figure \ref{fig:bp_filt_resid} shows BLASTPol $Q_{250}$, $Q_{250}^{\FSN}$, and the residual, with both the unfiltered and filtered maps. The structures visible towards the bottom of the residual maps in Figure \ref{fig:bp_filt_resid} are due to the signal from the Vela C molecular cloud and especially the \ion{H}{2} region RCW36, also seen in Figure 1 of \citet{Fissel16}. The agreement between filtered $Q_{250}$ and $Q_{250}^{\FSN}$ maps in Figure \ref{fig:bp_filt_resid}, as well as the absence of the 1/\textit{f} wings in the filtered residual maps, give us confidence that the high-pass filtering approach to removing systematic errors is both well-motivated and successful. 

Note that the scaling factor between the BLASTPol maps of a Stokes parameter and the corresponding FSN map at a given wavelength is equivalent to the value of the polarization spectrum at that wavelength, normalized at 850\,$\mu$m. We will calculate these scaling factors and thereby measure the polarization spectrum of a specific translucent molecular cloud in Section \ref{derive_polspec_sec}.

\subsection{Target cloud selection and properties} \label{target_props}
\begin{figure*}[ht]
\includegraphics[trim={0, 0.75in, 0, 0}, clip, width=0.95\textwidth, center]{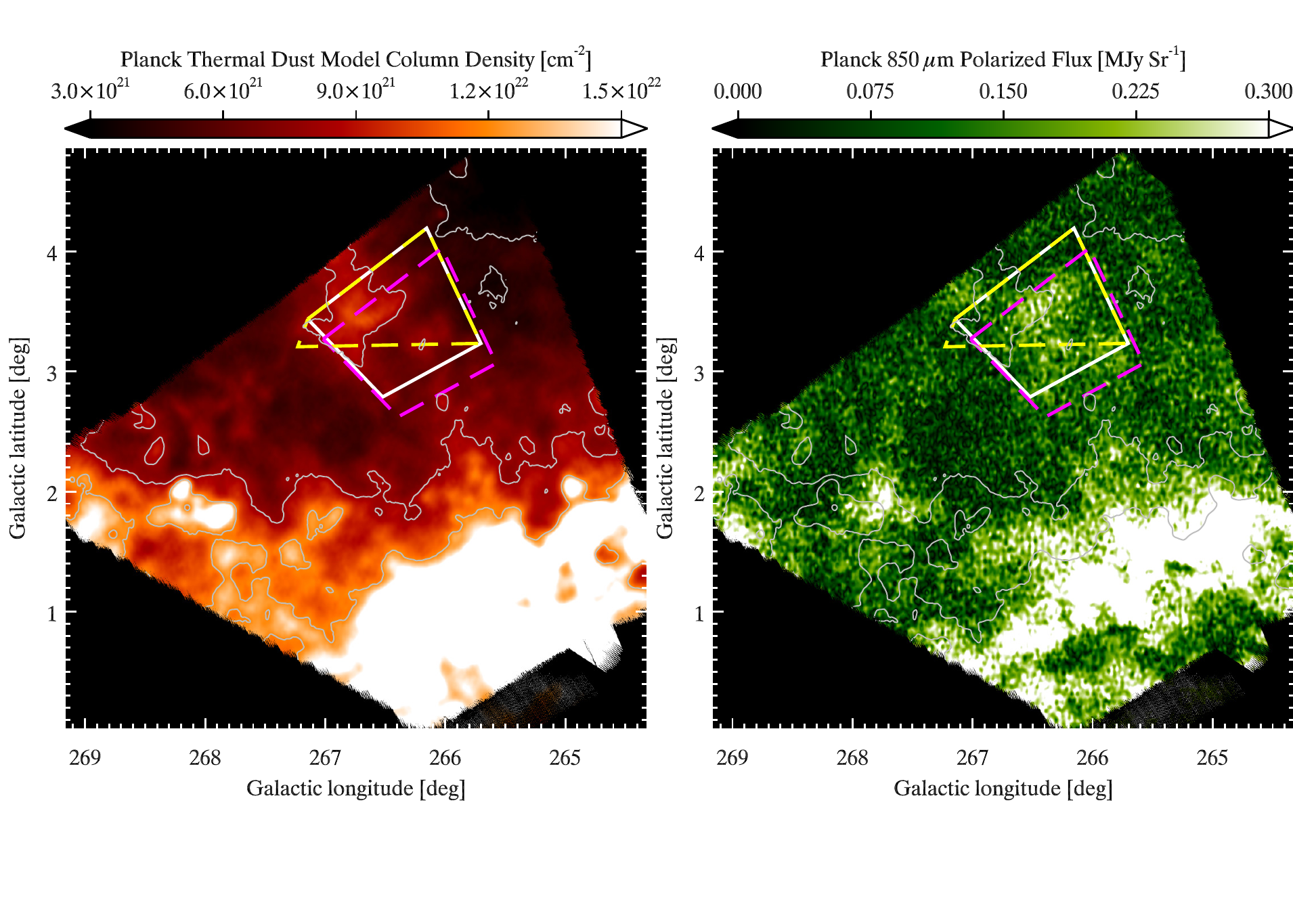}
\caption{\Planck thermal dust model column density map (left) and \Planck 850\,$\mu$m polarized flux (right) within the Vela Molecular Ridge region mapped by BLASTPol. The dust model map has been converted from submillimeter optical depth to H column density (see Section \ref{target_props}). Contours in both panels show column density levels at \{4, 8, 12\} x 10$^{21}$ cm$^{-2}$. In both panels, Region A is shown in white lines, Region B is shown in magenta dashed lines, and Region C is shown in yellow dashed lines. See Appendix \ref{appendix_sys} for discussion of Regions B and C.\label{fig:diff_reg}}
\end{figure*}

Having produced Stokes parameter maps that have been cleaned of their low-order spatial modes, we are in a position to choose a target cloud from the filtered set of maps. To select a target, we begin by considering the \Planck map of 850\,$\mu$m polarized flux ($P_{850} = \sqrt{Q_{850}^2 + U_{850}^2}$) within the region mapped by BLASTPol (Figure \ref{fig:diff_reg}, right). Four features stand out in this map; the Vela C molecular cloud is located near the southern edge, and three other isolated polarization structures are visible farther off the Galactic plane. Comparing with the \Planck thermal dust model column density map (Figure \ref{fig:diff_reg}, left), we can identify column density peaks near each of the isolated polarization structures, implying that the polarization signal is from a dust structure and not due to variation in the magnetic field inclination angle. Anticipating comparison with polarized dust models for the diffuse ISM, we choose to perform our analysis on the object with the lowest peak column density, which in this case is the northernmost polarization structure. The same type of analysis could be performed on the other object or a broader region of the BLASTPol map, but this is beyond the scope of the present work.

\begin{figure}
\includegraphics[trim={0.5in, 3in, 0.5in, 2.5in}, clip, width=1.0\columnwidth, center]{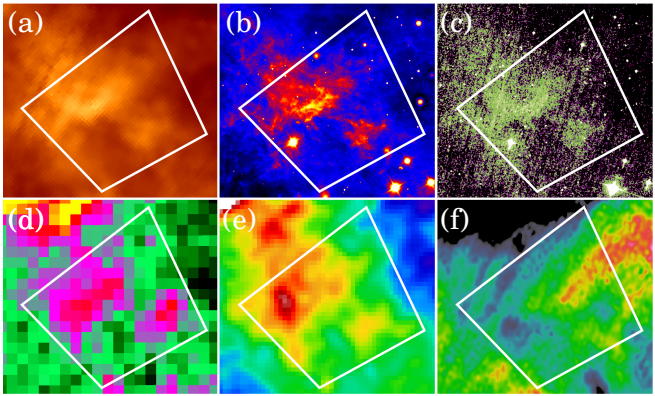}
\caption{From top left to bottom right, the target cloud as seen in IRIS 100\,$\mu$m, WISE 22\,$\mu$m, MSX 8\,$\mu$m, \Planck CO $J$ = 2$\to$1, stellar extinction from \citet{Dobashi05}, and \Planck thermal dust model temperature. Region A is shown in white in all panels. \label{fig:multi_wl}}
\end{figure}

The selected cloud is located at J2000 coordinates ($9^h03^m0^s.00$, $-43^{\circ}29^{\prime}00''$) (Gal. \textit{l} = $266.60^{\circ}$, b = $+3.46^{\circ}$). Structure related to the cloud can also be identified in maps of IRIS 100\,$\mu$m flux \citep{Miville-Deschenes05}, WISE 22\,$\mu$m flux \citep{Wright2010}, MSX 8\,$\mu$m flux \citep{Price01}, and \Planck CO \textit{J} = 2$\to$1 line emission (the highest spatial resolution \Planck CO data product; \citealt{planck2013-p03a}), as can be seen in Figure \ref{fig:multi_wl}. The presence of flux peaks in data sets covering a large spectral range gives us confidence that our target is a true physical object rather than the result of integrating unassociated emission along confused Galactic plane sightlines. We draw a quadrilateral around the target cloud that encompasses the peaks in both total intensity and polarized flux while avoiding the border of the BLASTPol scan area, where artifacts due to the spatial filtering may be present. This roughly 1 deg$^2$ quadrilateral is shown in white in Figure \ref{fig:diff_reg}, and we will refer to the region it encloses as ``Region A". Our results will be based on the analysis of signals within this region. In Appendix \ref{appendix_sys} we investigate the sensitivity of the result to perturbations to the map region under consideration. 

To better describe the target cloud's physical properties, we again consider the \Planck thermal dust model data products. The maximum 353\,GHz optical depth within Region A is $1.29 \times 10^{-4}$. Assuming a value for the 353\,GHz opacity ($\sigma_{353}$) of $12 \times 10^{-27}$\,cm$^2$ H$^{-1}$ \citep{planck2013-p06b}, the column density of H atoms is then $N_{\mathrm{H}} = \tau_{353}/\sigma_{353} = 1.1 \times 10^{22}$\,cm$^{-2}$. The true column density associated with the cloud is likely to be lower, though, as the target is very close to the Galactic plane, and emission from the entire line of sight will also contribute to the total column density. We roughly estimate the background column density from nearby lower-intensity regions at approximately the same galactic latitude in our maps. This background level is estimated to be $6 \times 10^{21}$\,cm$^{-2}$, leaving the target cloud with a column density of $5.0 \times 10^{21}$\,cm$^{-2}$. Converting this value to $A_V$ with a scaling factor $N_{\mathrm{H}}/A_V = 1.9 \times 10^{21}$\,cm$^{-2}$\,mag$^{-1}$ (\citealt{Bohlin78}; \citealt{Rachford09}), we find a maximum total extinction of 2.6\,mag, or under a crude approximation of being a spherical cloud, a center-to-edge $A_V$ of 1.3\,mag.

Furthermore, in the \Planck thermal dust model temperature map, we see a narrow distribution of temperatures within Region A (Figure \ref{fig:multi_wl}), centered around 18\,K with a range from 17.6\,K to 18.4\,K. Importantly, there is little if any correlation between the temperature map and the column density map. This fact implies that the interior of the target cloud is not strongly shielded from the external ISRF, so grain temperatures are not a strong function of location within the cloud. The mostly uniform temperature, together with the column density estimate above, and the presence of a peak in CO \textit{J} = 2$\to$1 emission (Figure \ref{fig:multi_wl}) all point to the target cloud having properties consistent with a translucent molecular cloud \citep{Snow06}.

\section{Deriving the Polarization Spectrum} \label{derive_polspec_sec}

\begin{table*}
\begin{center}
\caption{Average values of fit parameters and average uncertainties}
\begin{tabular}{cccccccc}
\hline
\hline
$\lambda$&$a_\lambda$&$\sigma_{a\lambda}$&$b_\lambda$&$\sigma_{b\lambda}$&$c_\lambda$&$\sigma_{c\lambda}$&$\sigma_{\mathrm{sys}}$\\
\hline
250 $\mu$m&0.997&0.089&$-$0.057&0.094&0.071&0.132&0.017\\
350 $\mu$m&1.057&0.100&$-$0.026&0.051&0.011&0.034&0.050\\
500 $\mu$m&0.889&0.078&$-$0.017&0.022&0.014&0.031&0.046\\
\hline
\end{tabular}
\end{center}
\label{table:mainres}
\end{table*}

\subsection{Linear Fitting Method}
Within Region A, we sample the filtered BLASTPol $Q_\lambda$ and $U_\lambda$ maps and the filtered \Planck $Q_\lambda^{\FSN}$ and $U_\lambda^{\FSN}$ in each of the three BLASTPol bands approximately once per beam\footnote[2]{For the remainder of this paper, unless otherwise noted, we will refer only to the filtered maps.}. At the common resolution of $4.8^\prime$, this amounts to sampling the value in the 10" sized pixels on a 29-pixel grid in both map dimensions. We form sets of ordered pairs ($Q_\lambda^{\FSN}$ , $Q_\lambda$) and ($U_\lambda^{\FSN}$ , $U_\lambda$) from the sampled maps at each wavelength, and employ the IDL method \texttt{MPFITEXY2}\footnote[3]{\url{https://github.com/williamsmj/mpfitexy}} \citep{Williams10} to perform linear fits. \texttt{MPFITEXY2} is a wrapper to \texttt{MPFIT} \citep{Markwardt_MPFIT} that iteratively minimizes a $\chi^2$ statistic, taking into account statistical errors in both the $x$ and $y$ variables. The errors in each coordinate for the $Q$ and $U$ relations come from sampling the relevant BLASTPol and \Planck covariance maps. Formally, \texttt{MPFITEXY2} requires statistically independent samples as its inputs. Due to the Gaussian kernel used in generating the maps, it is not possible to draw entirely independent samples. However, by sampling the maps at the scale of the resolution, the correlations between adjacent samples will only have a small effect. The statistical errors will be discussed in detail in the next section.

\begin{figure*}[ht]
\includegraphics[width=0.92\textwidth, center]{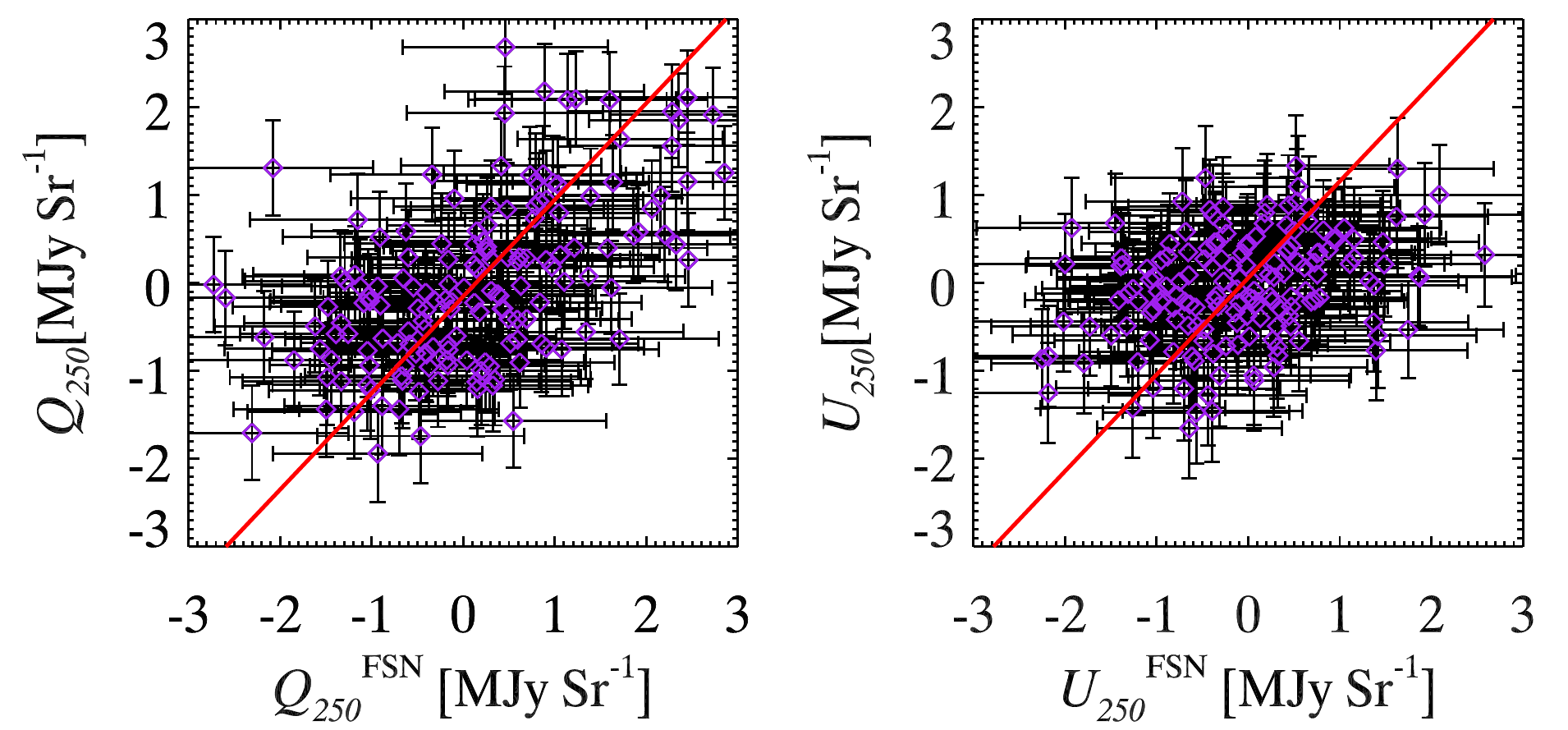}
\caption{Example scatter plots of filtered $Q_{250}$ vs. filtered $Q_{250}^{\FSN}$ (left) and filtered $U_{250}$ vs. filtered $U_{250}^{\FSN}$ (right) with statistically independent samples. Red lines show joint linear fit from \texttt{MPFITEXY2}.\label{fig:scatter}}
\end{figure*}

Importantly, \texttt{MPFITEXY2} fits a single slope to two data sets, with the possibility of two different offsets between them. In our application, the relation we wish to characterize is:
\begin{equation}\label{eq4}
Q_\lambda = a_\lambda Q_\lambda^{\FSN} + b_\lambda
\end{equation}
\begin{equation}\label{eq5}
U_\lambda = a_\lambda U_\lambda^{\FSN} + b_\lambda + c_\lambda
\end{equation}
where the best fit values of $a_\lambda$ give the 850\,$\mu$m normalized polarization spectrum in band $\lambda$. Sample $Q$ and $U$ scatter plots and the associated linear fit can be seen in Figure \ref{fig:scatter}. Because the polarization signal within Region A shows more contrast in $Q$ than in $U$ (because the magnetic field is nearly parallel to the galactic plane, and the Stokes parameters are referenced to Galactic coordinates), the $Q$ relation will tend to exert more influence on the fits. 

\texttt{MPFITEXY2} requires statistically independent samples in order to correctly determine the uncertainty in the resulting fit parameters. However Nyquist sampling, taking two samples per beam scale, effectively extracts all information from the map. Therefore to obtain final values for our fit parameters, we repeat the linear fits three more times, offsetting the set of sampled pixels by 2.4$^{\prime}$ in $l$, in $b$, and in both $l$ and $b$, respectively. Thus we obtain four different values, each with corresponding uncertainties, for each of the fit parameters in each band. In this way we can more completely sample the map area without artificially suppressing the statistical errors. We find the resulting parameters to be generally consistent across the four fits. In Table \ref{table:mainres}, for each band we report the arithmetic means of the fit parameters as well as the arithmetic means of the associated statistical errors, averaging over the four independent sampling fits.

The fact that the values of $b_\lambda$ and $c_\lambda$ in Table \ref{table:mainres} are consistent with zero gives us further confidence that the scaling between the polarization signals in two bands provides physical insight into the properties of the observed cloud. The statistical errors on the slopes $a_\lambda$ are found to be approximately 0.08-0.10. The values of $a_\lambda$ are consistent with unity, within these errors, indicating an approximately flat polarization spectrum. The following sections will discuss the relevant statistical and systematic errors on the result of these linear fits.

\subsection{Treatment of Statistical Errors} \label{stat_errors}
The 1-$\sigma$ statistical errors on the fit parameters returned by \texttt{MPFITEXY2} are included in Table \ref{table:mainres}. The statistical errors in BLASTPol $Q$ and $U$ are produced as outputs of the mapmaking software, \texttt{TOAST}. Strictly speaking, these errors correspond to the unfiltered maps. The errors in the \Planck FSN $Q$ and $U$ come from scaling the variances in the \Planck 850\,$\mu$m Stokes maps by the relevant ratio of intensities:
\begin{equation}\label{eq6}
\sigma_{Q, \lambda}^{\FSN} = \sigma_{Q850} \frac{I_\lambda^{\mathrm{mdl}}}{I_{850}}
\end{equation}
\begin{equation}\label{eq7}
\sigma_{U, \lambda}^{\FSN} = \sigma_{U850} \frac{I_\lambda^{\mathrm{mdl}}}{I_{850}}.
\end{equation}
This relatively simple scaling of variances, rather than a full propagation of errors for $Q_\lambda^{\FSN}$  and $U_\lambda^{\FSN}$, is equivalent to the assumption that fractional errors in $I_{850}$ and $I_\lambda^{\mathrm{mdl}}$ are small compared to those in $Q_{850}$ and $U_{850}$. This is likely to be the case when the absolute polarization degree is small. To check that this is the case, we consider the map of $p$ from the \Planck 850\,$\mu$m data set, correcting for a potential positive bias using the method of \citet{Wardle74}: $p_{db} = \sqrt{p^2 - \sigma_p^2}$ where $p_{db}$ and $p$ are the de-biased and measured values of the polarization degree, respectively, and $\sigma_p$ is the uncertainty in the measured value. We find that the median value of $p_{db}$ within Region A is approximately 3.2\%, so we consider the approximations for $\sigma_{Q, \lambda}^{\FSN}$ and $\sigma_{U, \lambda}^{\FSN}$ in Equations (\ref{eq6}) and (\ref{eq7}) to be valid.  

Reduced $\chi^2$ values for the individual fits to statistically independent samples range from 0.59 ($\chi^2$ = 190.8 with $N_{DOF}$ = 321) to 0.76 ($\chi^2$ = 244.5 with $N_{DOF}$ = 321). The fact that this value is less than unity implies that the errors in the BLASTPol and/or \Planck data that we used in the linear fits were overestimated. However in Appendix \ref{fitbias} we use simulated data to show that when performing fits with good signal-to-noise, the expected overestimation of the errors supplied to \texttt{MPFITTEXY2} will not significantly bias the fit slopes $a_\lambda$. 

\subsection{Treatment of Systematic Errors} \label{sys_errors}
The last column of Table \ref{table:mainres} lists the estimated systematic error on the fit slope $a_\lambda$ for each of the three BLASTPol bands. A full description of tests we performed to obtain these estimates is presented in Appendix \ref{other_errs}. To summarize, we perform three variation tests, recomputing the polarization spectrum after changing each of the two filter parameters ($k_0$ and $n$) and the map region sampled. The range over which each of these three choices is varied is intended to represent reasonable uncertainty in the optimal method of isolating the cloud polarization signal. For each variation, we find the maximum deviation of the fit slope, $a_{\lambda}$ with respect to the fiducial case ($k_0$ = 4, $n$ = 6, Region A), and we consider this deviation to be an estimate of the systematic error associated with the choice that was varied\footnote[4]{For the systematic error tests we Nyquist sample the Stokes parameter maps rather than using independent samples. We do this for simplicity, as we found that the resulting changes in the fitted slopes $a_\lambda$ are generally below 0.02, and because our systematic error tests make no use of the statistical errors from the fits.}. Then, the three error estimates are added in quadrature to form an effective $\sigma_{sys}$, shown in Table \ref{table:mainres}. We caution that, since systematic variations are not in general Gaussian distributed, it is formally incorrect to handle them as if they were. Rather, the systematic errors reported characterize the typical scale of uncertainty associated with reasonable variations to our analysis method. We find systematic errors to be smaller than the statistical errors across the three bands as can be seen by comparing the third and last columns of Table \ref{table:mainres}.

\subsection{Results}
\begin{figure}
\includegraphics[width=\columnwidth, right]{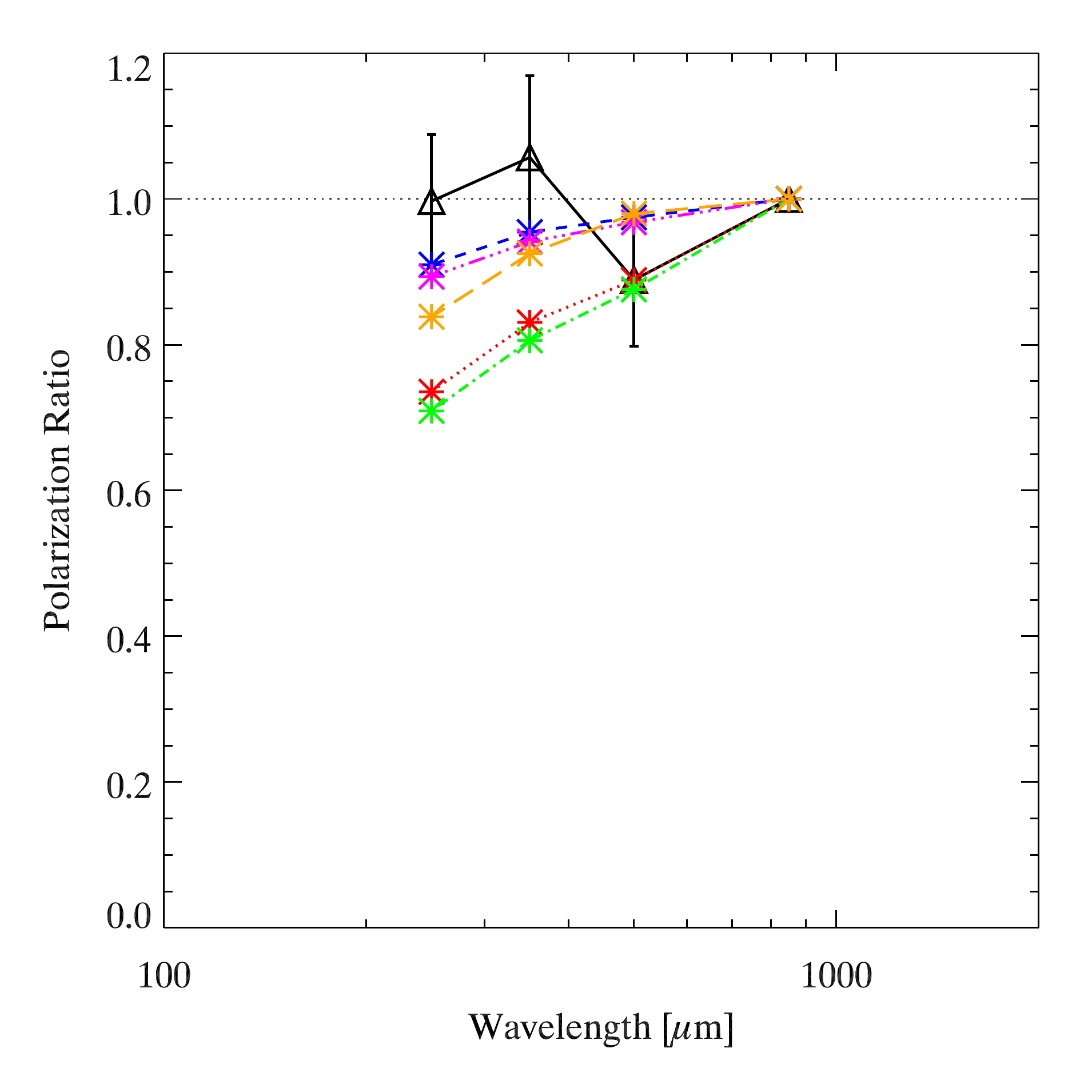}
\caption{Observed normalized polarization spectrum $p(\lambda)/p(850\,\mu m)$ (in black) along with \citet{DF09} models 1 through 4 (in red, blue, green, and pink, respectively). The \citet{Bethell07} polarization spectrum model is shown in orange. All models have been integrated over the four spectral bands and also normalized at 850\,$\mu$m.\label{fig:polspec_mdls}}
\end{figure}

Figure \ref{fig:polspec_mdls} shows our result for the normalized polarization spectrum in graphical form, plotted in black. The plotted error bars show the quadrature sum of the systematic and statistical errors on $a_{\lambda}$. We again caution that this is only for the purpose of visualizing the scale of variability. Formally, statistical and systematic errors should be handled separately, as in Table \ref{table:mainres}.

\section{Comparison With models}
\subsection{Models of Draine \& Fraisse (2009): the diffuse ISM} \label{DF09_model_sec}
Having measured the polarization spectrum of a translucent molecular cloud and constrained our uncertainties, we are now in a position to draw comparisons with theoretical predictions. Although no polarization spectrum models are available that specifically consider the conditions in a translucent molecular cloud, \citeauthor{DF09} (\citeyear{DF09}, hereafter DF09), produced four models of the polarization spectrum in the diffuse ISM. In these models, dust was simulated as a mixture of spheroidal silicate and graphite grains. In all of the DF09 models, the dust optical properties are specified via the prescription of \citet{Draine03} and the emission spectrum is calculated from temperature distributions following \citet{DL07}, but the axis ratio of the spheroids is varied among the models. Importantly, in two of the four models (their models 1 and 3), graphite grains are taken to be spheres and thus incapable of producing polarized emission due to magnetic alignment, while the silicate grains are taken to be oblate spheroids and potentially aligned depending on their size. In the other two models (their models 2 and 4), both silicate and graphite grains are oblate. For each of the four models, the dust abundance and alignment efficiency are set as a function of grain size in order to agree with observations of total and polarized extinction at visible and near-infrared (NIR) wavelengths. Then having set the properties of the dust population in each of the four scenarios, the total and polarized emission is computed assuming the dust is heated by an ISRF similar to that in the local solar neighborhood. The DF09 models, integrated over the four bands and normalized at 850\,$\mu$m, are plotted in red, blue, green, and pink in Figure~\ref{fig:polspec_mdls}, with our observed polarization spectrum in black. Looking at the polarization spectra, the four DF09 models sort into two classes; the normalized polarization spectra for models with spherical, unaligned graphite grains fall by a factor of approximately 0.7 at 250\,$\mu$m compared to 850\,$\mu$m, whereas the models with both silicate and graphite grains aligned have flatter normalized polarization spectra, falling by only about a factor of 0.9 over the same wavelength range. Our observational results, then, would seem to conflict with the models employing unaligned graphite grains. We will return to this point later in the paper.

\subsection{Underlying physics} \label{underlyingPhys}

In the diffuse ISM, the ISRF is largely unattenuated, and every dust grain is exposed to an identical radiation field. In this case, dust temperature and alignment efficiency will depend only on grain properties such as size, shape, and composition, and not on location. According to the dust model by \citet{LD01}, larger grains (both silicate and graphite) will tend to be cooler than their smaller counterparts due to their relatively higher absorption (and emission) efficiency \citep{Draine11}. 

Additionally, grain alignment efficiency is known to be correlated with grain size from optical and NIR polarimetry observations \citep{Kim95}, with larger grains tending to be better aligned on average than smaller ones. This correlation matches the prediction from the radiative alignment torques (RATs) picture of grain alignment, in which the radiation field is responsible for aligning grains with respect to their local magnetic field (\citealt{Dolginov76}; \citealt{LH07}; \citealt{Andersson15} and references therein). Note that although the DF09 models do not invoke the RATs mechanism explicitly, the dependence of alignment efficiency derived by modeling the wavelength dependence of optical/NIR polarimetric observations produces a RATs-like correlation between grain alignment and size (i.e.\ one where larger grains are better aligned). 

Thus in the diffuse ISM, larger dust grains tend to be colder than average and better-aligned than average. Accordingly the DF09 polarization spectrum models all have positive slopes in the submillimeter, as thermal emission from aligned grains becomes relatively more significant with increasing wavelength. Furthermore, the DF09 models with unaligned graphite grains (modeled as a shape effect) have steeper slopes; this is due to silicate grains being colder than equivalent-size graphite grains as well as the relatively higher opacity of silicate compared to graphite grains at longer wavelengths. Both of these factors give more weight to the (partially polarized) emission from aligned silicate grains at longer wavelengths.

\subsection{Bethell et al. (2007): dense molecular clouds} \label{B07_model_sec}
The DF09 models were built to simulate conditions in the diffuse ISM, whereas our observed polarization spectrum is from a molecular cloud, albeit one with relatively low column density. Within these clouds, one might expect a difference in the penetrating radiation field (which contributes to determining the total and polarized dust emission) compared to the radiation field in the diffuse ISM. Thus, in order for observations of the polarization spectrum in a translucent molecular cloud to be relevant for discriminating among the diffuse ISM models, it is first necessary to confirm that the conditions in the two environments are similar with regard to the factors that would affect the polarization spectrum shape.

Details of the relationship between extinction, grain alignment, and dust temperature were thoroughly studied for a dense molecular cloud ($A_V$ = 10\,mag, center-to-edge) by \citeauthor{Bethell07} (\citeyear{Bethell07}, hereafter B07). There, the authors simulated magnetohydrodynamic turbulence and radiative transfer of the external ISRF within the clumpy, porous cloud. They use the dust optical properties for the silicate-graphite mixture described by \citet{DL84} and the power law grain size distribution of \citeauthor{MRN77} (\citeyear{MRN77}, hereafter MRN77). Alignment is imposed on silicate grains only, based on an empirical prescription for the RATs mechanism from \citet{CL05} that relates alignment efficiency with visual extinction and gas number density. Under these conditions, B07 find a polarization spectrum for a dense molecular cloud that is largely flat through submillimeter wavelengths longer than about 200\,$\mu$m while falling sharply toward shorter wavelengths than that. Notably, \citet{Gandilo16} found this model to be consistent with BLASTPol and \Planck observations of the 250\,$\mu$m to 850\,$\mu$m polarization spectrum in the Vela C dense molecular cloud. The orange trace in Figure \ref{fig:polspec_mdls} shows the B07 polarization spectrum model integrated over the four bands and normalized at 850\,$\mu$m.

In a cloud with some significant extinction, dust in denser regions of the cloud will be exposed to a radiation field that is redder and dimmer than that seen by dust near the cloud surface. Since, in the RATs paradigm, the radiation field is responsible for aligning grains as well as heating them, the dust in the interior will be on average cooler and less well-aligned than the typical dust in the cloud as a whole. This extinction-temperature-alignment correlation (ETAC), whereby warmer surface dust is better aligned than cooler interior dust, runs counter to the correlation described in Section \ref{underlyingPhys} for the diffuse ISM. The result is that when considering a dense cloud in total, some aligned grains exist at temperatures both above the average dust temperature in the cloud (because they are better illuminated) and below it (because they are larger). Thus the average temperature of aligned grains is closer to the average temperature of all grains, and the resulting polarization spectrum is flatter than would be seen without the ETAC effect. 

\subsection{Relevance to translucent clouds} \label{transcloud_relevance}

The core issue for the present work, then, is the source of the discrepancy between the available models and how it would apply to our observed translucent cloud. Both DF09 and B07 show models in which only silicate (and not graphite) grains are aligned, but the former find a steeply rising polarization spectrum while the latter find it to be flatter. To some degree this difference in shape is due to the fact that the DF09 models employ a prescription for the dust optical properties that assumes different values for the spectral index $\beta$ for silicate and graphite grains (1.6 versus 2.0, respectively), whereas the dust in the B07 model has $\beta$ = 2.0 for both materials. Additionally however, B07 show that temperature distributions for dust grains of a given size and material in their model typically have widths of several K. This temperature variation is a result of the variation in shielding within the dense cloud and is manifested as a flatter polarization spectrum due to the ETAC effect described above.

Because dust temperature variations are likely to significantly affect emission near the peak wavelength of thermal emission, it is reasonable to assume the ETAC effect is at least partially responsible for the flatter submillimeter polarization spectrum found by B07. In this case, the translucent cloud that we have observed is likely to lie somewhere between these two extreme cases of low- and high-extinction modeled by DF09 and B07, respectively. For our current discussion, we seek to determine whether dust like that simulated by DF09 would emit with a significantly flatter polarization spectrum due to the ETAC effect in a translucent molecular cloud. In order to understand which model is a more appropriate comparison for our data, we must understand the point at which the ETAC effect becomes important as column density increases.  

To this end, we can gain insight into the onset of extinction effects on grain alignment from Figure 14 in B07. There, the authors quantify the disagreement between the direction of the mass-weighted, plane-of-sky projection of the magnetic field in their simulated cloud and the direction inferred from synthetic observations of the cloud's polarization signal. These data are shown in a histogram binned by column density, where the authors find an increasing discrepancy between the polarization-inferred and true magnetic field angles with increasing column density. However, on its own this divergence could be due to either a decrease in efficiency of alignment mechanisms or a projection effect from variations in field angle correlated with dust temperature along the line of sight in the model.

In order to separate these two possibilities, B07 also show the corresponding data in a version of their simulation where all grains are forced to be perfectly aligned to the local magnetic field. The column density at which the magnetic field angle agreement in the realistic simulation deviates from the case of forced alignment represents the point where the shielding of the cloud becomes important, corresponding to the onset of the ETAC effect. 

In the B07 simulated cloud, this deviation from the perfect alignment case occurs at a column density $N_{\mathrm{H}} \simeq 4 \times 10^{21}\,$cm$^{-2}$. For comparison, in our target cloud for the adopted opacity, the observed peak column density is approximately $5 \times 10^{21}$\,cm$^{-2}$. Thus while there might be some loss of grain alignment in the center of the target cloud, most of the cloud volume would have a radiative environment like the diffuse ISM and thus will not be subject to the ETAC effect. This picture is reinforced by the fact that the \Planck thermal dust model temperature is largely constant to within approximately 0.5\,K across Region A (see Section \ref{target_props}), indicating that the ISRF penetrates the observed cloud for the most part unattenuated. Based on these comparisons of column density and temperature, we conclude that comparison between our observed polarization spectrum and the models of DF09 is justified, and we further conclude that there exists a significant discrepancy between the observation and the DF09 models in which only silicate grains have the potential to be aligned.

Generalizing this result to highlight the underlying physics, we see that BLASTPol observations cover a wavelength range that makes it possible to distinguish between physical dust models. In particular, these observations present a challenge to models in which dust alignment degree and grain temperature are strongly correlated (e.g.\ due to the presence of major components with distinct temperature and alignment distributions).  

Finally, we note that even the DF09 models 2 and 4, in which graphite grains are partially aligned, disagree with our observed polarization spectrum at approximately the 1-$\sigma$ level. We do not claim to see a significant discrepancy with these models, and even if the ETAC effect were active in the observed cloud, it would also tend to flatten the polarization spectra of these models relative to the diffuse ISM. However, as we have already seen multiple different physical models that produce polarization spectra that are largely flat in the submillimeter (i.e.\ DF09 models 2 and 4 and the B07 model), it should be clear that a flat polarization spectrum does not uniquely identify a single set of physical conditions. Thus the remainder of this paper will focus on verifying the disagreement with DF09 models 1 and 3 rather than affirming models 2 and 4.


\section{Simple Analytic Models for the ETAC} \label{toymdls}
\subsection{Motivation and general method}\label{mdls}
In this section, we use simplified analytic models of grain population alignment and emission in order to validate our understanding of the relative importance of the factors influencing polarization spectra. In particular, in Section \ref{DFmdl} we aim to confirm our earlier conclusion that the DF09 models 1 and 3 have strong positive slopes due to a combination of the aligned grains being colder on average than the unaligned grains and the difference between $\beta$ for silicate and graphite grains. In Section \ref{B07mdl} we test our conclusion that the flatter spectrum seen by B07 is plausibly due to the ETAC effect. Finally, in Section \ref{obscloud_mdl} we validate our finding that the ETAC effect is relatively unimportant in the translucent molecular cloud observed with BLASTPol. This confirmation will serve to reinforce our previous conclusion that the polarization spectrum we have observed disfavors DF09 models 1 and 3.

To these ends, we construct models for polarization spectra under several simplifying assumptions. Generally, the polarization spectrum $p(\lambda)$ is the ratio of polarized to total emission: $p(\lambda) \equiv P(\lambda)/I(\lambda)$. To estimate $P(\lambda)$ and $I(\lambda)$, we divide the total dust population into three subsets: all silicate grains, all graphite grains, and aligned silicate grains. The unpolarized emission from grains of a given material is calculated to be:
\begin{equation} \label{eq10}
I_m(\lambda) \propto \int_{a_{\mathrm{min}, m}}^{a_{\mathrm{max}, m}} \Big(\frac{dn}{da}\Big)_m\, C_{\mathrm{abs}}(a, \lambda, m)\, B_{\lambda}(T_{\mathrm{eq}}(a, m))\, da,
\end{equation}
where \textit{m} indexes the material, either `s' or `g' for silicate or graphite grains, respectively. We take $dn/da$ from the MRN77 grain size distribution, for which the number of grains of a given effective radius ($a$, the radius of an equivalent-volume sphere) scales as $a^{-3.5}$ within the range [$a_{\mathrm{min}}, a_{\mathrm{max}}$], which can vary depending on grain material. $C_{\mathrm{abs}}$, the dust absorption cross section, depends on grain size, material, and wavelength. Following \citet{Draine11}, we take
\begin{equation} \label{eq11}
C_{\mathrm{abs}}(a, \lambda, m) = \pi a^2 Q_{0, m} \Big(\frac{a}{1\,\mu m}\Big) \Big(\frac{\lambda}{100\,\mu m}\Big)^{-\beta_m}
\end{equation}
where $Q_{0, m}$ depends on the grain material, and we allow for different materials to have different spectral indices ($\beta_m$). For silicate and graphite grains, $Q_0$ is $1.4 \times 10^{-2}$ and $1.0 \times 10^{-2}$, respectively. We assume that each grain exists at its equilibrium temperature $T_{\mathrm{eq}}$ which is determined only by its size and material. Note that Equation (\ref{eq11}) effectively treats all grains as spheres for the purposes of calculating total emission; we treat $C_{\mathrm{abs}}$ as independent of any grain body or radiation coordinate system. To summarize, the total unpolarized dust emission $I(\lambda)$ is given by:
\begin{align} \label{eq12}
I(\lambda) \propto &\sum_m \Big[\pi Q_{0,m}  \Big(\frac{\lambda}{100 \mu m}\Big)^{-\beta_m} \times \\
&\int_{a_{\mathrm{min}, m}}^{a_{\mathrm{max}, m}} a^{-0.5}\,B_\lambda(T_{\mathrm{eq}}(a, m))\,da\Big]. \nonumber
\end{align}

We note that this construction assumes an equal number of silicate and graphite grains at each grain size. A more complete treatment would apply weighting factors ($A_m$) to the contributions of silicate and graphite grains in Equation (\ref{eq12}). Estimates of the ratio of silicate volume to graphite volume have found this value to be an order-unity factor \citep{WD01}. Rather than enforcing this ratio in our simple models, we note that in the MRN77 distribution, the total volume for grains of a given material is proportional to $A_m \sqrt{a_{\mathrm{max}}}$ if $a_{\mathrm{max}} \gg a_{\mathrm{min}}$. Thus if $a_{\mathrm{max}}$ is on the same order for silicate and graphite grains in our models, the ratio of total volumes will roughly match observations, and we can neglect the weighting factors $A_m$ as a crude approximation. 

For the polarized emission $P(\lambda)$, we assume that all silicate grains larger than some size $a_{\mathrm{align}}$ are aligned. The expression for $P(\lambda)$ is then:
\begin{align} \label{eq13}
P(\lambda) \propto & \Big[\pi Q_{0,\mathrm{s}}  \Big(\frac{\lambda}{100 \mu m}\Big)^{-\beta_\mathrm{s}} \times \\
&\int_{a_{\mathrm{align}}}^{a_{\mathrm{max}, \mathrm{s}}} a^{-0.5}\,B_\lambda(T_{\mathrm{eq}}(a, \mathrm{s}))\,da \Big]. \nonumber
\end{align}
Note that by relying on the expression for $C_{\mathrm{abs}}$ that assumes spherical grains, we are effectively assuming that the difference in absorption cross sections for radiation polarized parallel and perpendicular to the grain's alignment axis is proportional to $C_{\mathrm{abs}}$. The polarization spectrum $p(\lambda)$ is then calculated as the ratio of polarized to total emission and normalized to its value at 850\,$\mu$m. Thus, in order to calculate the polarization spectrum in this simple model, it is necessary to specify $a_{\mathrm{min}}$, $a_{\mathrm{max}}$, and $\beta$ for graphite and silicate grains, $a_{\mathrm{align}}$ for aligned silicate grains, and the temperature distributions $T_{\mathrm{eq}}(a, m)$.

\subsection{Reproducing Draine \& Fraisse (2009)}\label{DFmdl}
We begin by attempting to reproduce the DF09 models 1 and 3 in which only silicate grains are capable of alignment. For the temperature function $T_{\mathrm{eq}}(a, m)$ in Equations (\ref{eq12}) and (\ref{eq13}), we employ the the functions of \citet{LD01}, which give grain temperature as a function of grain material, radiation field intensity, and grain effective radius under similar assumptions for the dust optical properties as were used in the DF09 models. For silicate and graphite grains, we assume the functions $T(a)$ given their parameter $\chi_{\mathrm{MMP}}$ = 1 (approximating the local solar ISRF). To match the dust optical properties used in the DF09 models, we set $\beta_{\mathrm{s}}$ = 1.6 and $\beta_{\mathrm{g}}$ = 2.0. For each grain material, we set $a_{\mathrm{max}}$ at the point where the DF09 model 1 distribution of dust mass per logarithmic interval reaches half its maximum value. Based on this criterion, we find $a_{\mathrm{max}}$ to be 0.44\,$\mu$m for silicate grains and 0.13\,$\mu$m for graphite grains. We set the minimum grain size $a_{\mathrm{min}}$ to be 0.01\,$\mu$m for both materials, noting that this is significantly below the peak of the mass distribution for both. Finally, $a_{\mathrm{align}}$ is taken from DF09 to be a constant value 0.1\,$\mu$m based on the sharp rise they find in alignment fraction as a function of grain size. 

\begin{figure}
\includegraphics[width=\columnwidth, right]{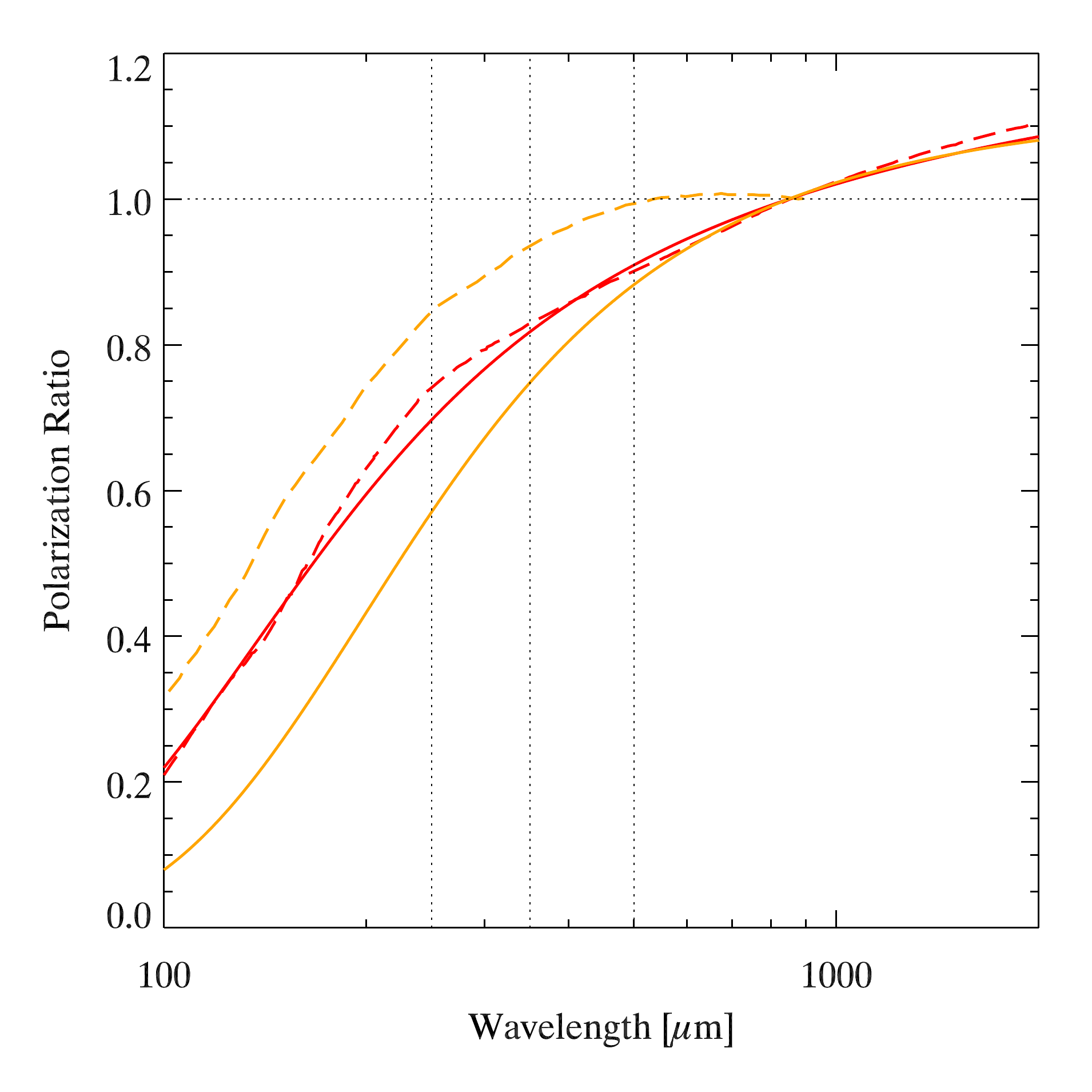}
\caption{Normalized polarization spectrum models. The simple analytic model described in Section \ref{DFmdl} is shown in solid red, compared to Model 1 from DF09 in dashed red. The simple analytic model described in Section \ref{B07mdl} is shown in solid orange, and the model produced by \citet{Bethell07} is shown in dashed orange. Vertical dotted black lines show the central wavelengths of the three BLASTPol bands.}\label{fig:df_beth_mdls}
\end{figure}

The polarization spectrum, normalized at 850\,$\mu$m is shown in red in Figure \ref{fig:df_beth_mdls} along with the DF09 model~1. We note the close agreement between the two. Comparing the spectra at the three BLASTPol band centers, the largest discrepancy is at the 5\% level in the 250\,$\mu$m band. This agreement implies that our simplifying assumptions effectively capture the basic physics treated in detail in the DF09 models.

\subsection{Reproducing Bethell et al. (2007)}\label{B07mdl}
Now we employ the same method described in Section \ref{mdls}, altering the temperature distributions and relevant grain parameters to match the conditions simulated by B07 within a dense molecular cloud. In this case, we take dust temperature distributions from work by \citet{Bethell04}, who performed a simulation of radiative transfer within a turbulent molecular cloud similar to that employed by B07. Those authors reported average dust temperatures for both silicate and graphite grains at $a$ = 0.001, 0.01, 0.1 and 1\,$\mu$m. To estimate dust temperatures as a function of grain size in the B07 model, we linearly interpolate the \citet{Bethell04} temperature values as a function of log$_{10}(a)$. These functions $T(a)$ are then inserted into Equations (\ref{eq12}) and (\ref{eq13}) to compute the total and polarized emission. For the grain size distributions, B07 use MRN77 distributions, and they specify their $a_{\mathrm{min}}$ and $a_{\mathrm{max}}$ to be 0.005\,$\mu$m and 0.5\,$\mu$m for both silicate and graphite grains. Additionally, we take the dust spectral indices $\beta_{\mathrm{s}} = \beta_{\mathrm{g}} = 2.0$, as was the case in the B07 model. Finally, it is necessary to set the value of $a_{\mathrm{align}}$. In the B07 simulated cloud, the size of the smallest aligned grain varies significantly with extinction and gas number density within turbulent clumps. For our purposes in constructing simple models, we take $a_{\mathrm{align}}$ = 0.3\,$\mu$m, as we expect the value to be larger than was the case in the diffuse ISM but still less than $a_{\mathrm{max}}$ in order for our model to produce any polarized signal. Since the average temperature of silicate grains in the \citet{Bethell04} model does not change much with grain size, our model is not very sensitive to the choice of $a_{\mathrm{align}}$. We verify this by repeating the procedure with $a_{\mathrm{align}}$ = 0.2\,$\mu$m and 0.45\,$\mu$m, and we see that the resulting normalized polarization spectrum changes by \textless 1\% across the submillimeter wavelengths of interest.

The resulting normalized polarization spectrum is shown in orange along with the B07 model in Figure \ref{fig:df_beth_mdls}. Here we see significant disagreement between our simple analytic model and the detailed simulation, approaching 30\% at 250\,$\mu$m. Evidently in the case of the dense molecular cloud, our assumptions of a single value of $a_{\mathrm{align}}$ and a dust temperature depending only on grain size and material fail to reproduce the largely flat polarization spectrum (for $\lambda \gtrsim$ 200 $\mu$m) of B07. Those authors note this fact explicitly, explaining that the temperature of the smaller grains depends most on the blue end of the ISRF, which experiences the most significant variations within the cloud's clumpy structure. The result is that B07 see relatively broad and correlated distributions of $a_{\mathrm{align}}$ values and grain temperatures. Given our success in reproducing the polarization spectra corresponding to DF09 models 1 and 3 (see Section \ref{DFmdl}), it seems plausible that our failure to do the same for B07 is due mainly to the ETAC effect, as we have not attempted to model the ETAC-induced flattening of the polarization spectrum that seems likely to be present in B07 (See Section \ref{B07_model_sec}). A more detailed evaluation of the factors affecting the B07 model polarization spectrum is beyond the scope of the present paper. 

\subsection{Modeling the observed translucent cloud}\label{obscloud_mdl}
We now investigate whether the ETAC effect expected in a translucent cloud could cause dust like that in DF09 to produce a polarization spectrum consistent with our observations. Whereas in the previous sections we treated the dust emission with a single distribution of $T_{\mathrm{eq}}$ and a single value of $a_{\mathrm{align}}$, to crudely capture the ETAC effect we instead treat the translucent cloud as the sum of two spherically symmetric, equal-mass components: a relatively more shielded ``bulk'' and a relatively less shielded ``surface''. Because dust in the bulk will see a relatively dimmer and redder radiation field than dust in the surface, we will need to specify the grain alignment and temperature distributions for the surface and bulk separately.

Furthermore, the expressions for total and polarized emission will be modified from Equations (\ref{eq12}) and (\ref{eq13}) to include emission from both cloud components, so we calculate:
\begin{align}\label{eq14}
I(\lambda) \propto &\sum_i \sum_m \Big[\pi Q_{0,m}  \Big(\frac{\lambda}{100 \mu m}\Big)^{-\beta_m} \times \\ 
&\int_{a_{\mathrm{min}, m}}^{a_{\mathrm{max}, m}} a^{-0.5}\,B_\lambda(T_{\mathrm{eq}}^{(i)}(a, m))\,da \Big], \nonumber
\end{align}
\begin{align}\label{eq15}
P(\lambda) \propto &\sum_i \Big[\pi Q_{0,s} \Big(\frac{\lambda}{100 \mu m}\Big)^{-\beta_\mathrm{s}} \times \\ 
&\int_{a_{\mathrm{align}, i}}^{a_{\mathrm{max}, \mathrm{s}}} a^{-0.5}\,B_\lambda(T_{\mathrm{eq}}^{(i)}(a, \mathrm{s}))\,da \Big], \nonumber
\end{align}
where $i$ = 1 or 2 indexes the contribution from the surface and the bulk of the model cloud.

In order to split the spherical cloud of radius $R$ into two equal-mass (and equal-volume) components, we define a boundary surface at radius $r_0 = R/[2^{1/3}] \approx 0.79R$. We estimate that the extinction at radius $r$ depends on the shortest distance from $r$ to the cloud surface, reaching a maximum value of 1.3\,mag at $r = 0$ so that $A_V(r)$ = (1.3\,mag$)(1 - r/R)$. Finally, we set characteristic values for the extinction in each component as the $A_V$ at the radius for which the component has equal mass inside and outside that radius\footnote[5]{Put another way, the bulk and surface $A_V$ take the values at the surfaces that enclose 25\% and 75\% of the total cloud's mass.}. The result of this calculation is that the extinction in the surface and bulk components will be represented by $A_V^{\mathrm{s}}$ = 0.12\,mag and $A_V^{\mathrm{b}}$ = 0.48\,mag, respectively.

\begin{figure}
\includegraphics[width=\columnwidth, right]{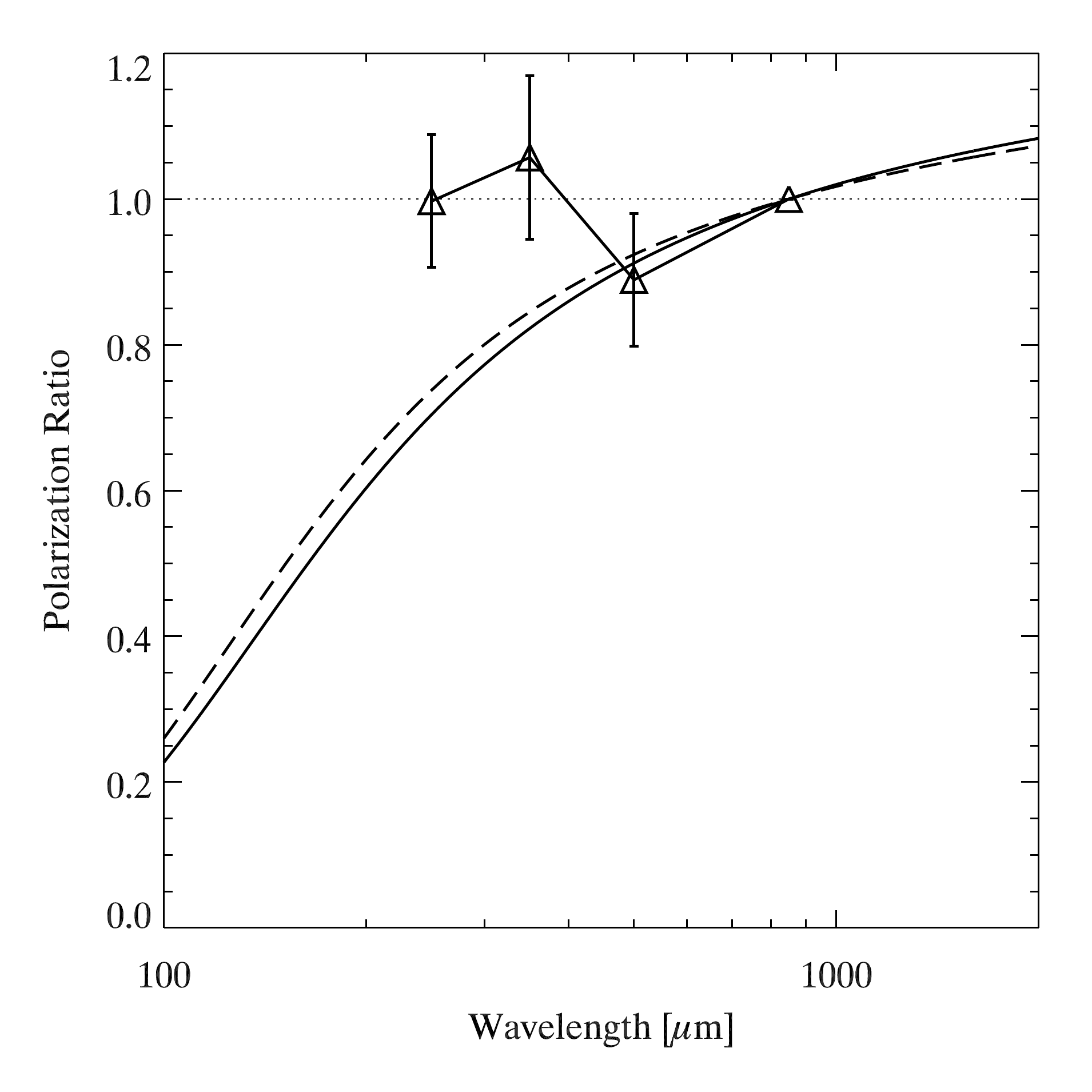}
\caption{Normalized polarization spectra from the surface/bulk two-component model (solid line) and the surface/bulk-max model (dashed line).The observed polarization spectrum data are shown with triangular symbols and error bars as in Figure \ref{fig:polspec_mdls}.} \label{fig:2_comp_mdls}
\end{figure}

To calculate the polarization spectrum in the two-component model, we set $a_{\mathrm{min}}$ and $a_{\mathrm{max}}$ for silicate and graphite grains to the same values used in the diffuse ISM model in Section \ref{DFmdl} and the same in the surface and the bulk. We determine $a_{\mathrm{align}}$ and modify the \citet{LD01} temperature distributions based on the extinction in each component. The details of these calculations are described in Appendix \ref{appendix_2compmdl}, and the parameters used in the model are summarized in Table \ref{2compparams}. The resulting normalized polarization spectrum is shown in Figure \ref{fig:2_comp_mdls}. 

\begin{table*}[ht]\label{2compparams}
\begin{center}
\caption{Two-component analytic model parameters}
\begin{tabular}{cccc}
\hline
\hline
Parameter&Surface&Bulk&Bulk-max\\
\hline
Effective $A_V$ [mag]&0.12&0.48&1.3\\
$a_{\mathrm{min}, \mathrm{s}}$ [$\mu$m]&0.01&0.01&0.01\\
$a_{\mathrm{max}, \mathrm{s}}$ [$\mu$m]&0.44&0.44&0.44\\
$a_{\mathrm{min}, \mathrm{g}}$ [$\mu$m]&0.01&0.01&0.01\\
$a_{\mathrm{max}, \mathrm{g}}$ [$\mu$m]&0.13&0.13&0.13\\
$a_{\mathrm{align}}$ [$\mu$m]&0.1&0.2&0.4\\
$\beta_\mathrm{s}$&1.6&1.6&1.6\\
$\beta_\mathrm{g}$&2.0&2.0&2.0\\
\hline

\end{tabular}
\end{center}
\end{table*}

Although the two-component model produces a normalized polarization spectrum that is slightly flatter than our simple models attempting to reproduce the DF09 results ($<$ 1\% difference at 250\,$\mu$m, comparing the solid line in Figure \ref{fig:2_comp_mdls} to the solid red line in Figure \ref{fig:df_beth_mdls}), it remains significantly steeper than the polarization spectrum observed by BLASTPol. To investigate whether the rising spectrum we find is due to an underestimated shielding of the ISRF in the bulk, we repeat the above analysis instead treating the bulk as if it were shielded by the maximum center-to-edge $A_V$ seen at the cloud center, i.e.\ 1.3\,mag. The resulting parameters used in this ``bulk-max" model are listed in the last column of Table \ref{2compparams}. We calculate the modified polarization spectrum by combining the emission from the surface with the bulk-max component, and the result is also shown in Figure \ref{fig:2_comp_mdls}. We note that the normalized polarization spectra resulting from these simple models are very little changed from the polarization spectrum we found in Section \ref{DFmdl} which did not include the ETAC effect. Thus we conclude that the ETAC effect is very weak in translucent molecular clouds. We see that the surface/bulk-max modification does indeed further flatten the polarization spectrum relative to the DF09 result and to the surface/bulk model, but not to the degree that would be necessary to agree with observations. 

\section{Discussion and Summary}

The simple analytic models presented in Section \ref{toymdls} are meant to qualitatively investigate the impact of varying environmental parameters on the polarization spectrum. By employing several plausible simplifications, we conclude the following:

\begin{enumerate}
\item The DF09 models for which polarized emission only arises from aligned silicate grains are well-fit by a model where all silicate grains larger than a certain size are perfectly aligned, and dust temperature can be specified by grain size and material only. In this case we note that the rising polarization spectrum with wavelength is a consequence of the difference between the properties of silicate and graphite grains. Graphite grains are overall warmer, and their emission falls off more steeply with increasing wavelength, leading to relatively more (unpolarized) emission from unaligned graphite grains at shorter wavelengths.

\item In attempting to reproduce the B07 model, the same method of simplification fails to generate the flat polarization spectrum found via detailed simulation. Instead we see a strongly rising spectrum, which can be attributed to the same temperature difference between silicate and graphite grains previously mentioned. The difference between our simple model and the detailed treatment of B07 is likely the failure of our assumption that a dust grain's temperature can be determined by its size and material only. In the B07 model, grains of a given size have a broad distribution of temperatures, owing to the variations in shielding of the ISRF within the turbulent cloud. Furthermore, the warmest silicate grains of a given size are the least shielded, and thus most likely to be aligned. Therefore it seems that the average temperature of aligned grains is warmer than the average temperature of all silicate grains, and closer to the average temperature of all grains (silicate and graphite). This is the basis of the ETAC effect, which our simple model neglects.

\item When we attempt to explain the observed translucent cloud with a simple implementation of the ETAC effect, we find that the effect is not strong enough to explain the observed flat polarization spectrum given dust properties similar to DF09 models 1 and 3 and plausible environmental parameters. This finding affirms our earlier conclusion in Section \ref{transcloud_relevance} that increased shielding is likely not responsible for the flat polarization spectrum as seems to be the case in the B07 simulated dense molecular cloud. Thus comparison to polarized models of the diffuse ISM is appropriate, and on this basis we note a significant disagreement between our observations and the DF09 models where all polarization is due to aligned silicate (and not graphite) dust grains.
\end{enumerate}   

Though the present work disfavors the DF09 models with only silicate grains capable of alignment, we emphasize that this result does not necessarily confirm the other DF09 models where both silicate and graphite grains may be aligned. Indeed, these models would seem to be in conflict with the non-detection of polarization in the 3.4\,$\mu$m C-H stretch mode absorption feature \citep{Chiar06}.

The observed submillimeter polarization spectrum presented here is complementary to the finding of \citet{planck2014-XXII}, which saw a diffuse ISM polarization spectrum which is flat or decreasing with wavelength from 850\,$\mu$m to 4.3\,mm. Those authors noted the disagreement with polarization spectrum models predicting a rising polarization degree with wavelength, and suggested that the discrepancy may be due to variation between the dust emission spectral index $\beta$ for silicate and carbonaceous grains. At the submillimeter wavelengths observed by BLASTPol, differences in the temperatures of the grain populations strongly affect the polarization spectrum. This is not true for the longer wavelengths observed by \textit{Planck}, where differences in $\beta$ will become more important.

Our present work is also adjacent to recent studies of spectral and spatial variation of the dust polarization signal (\citealt{planck2014-XXX}; \citealt{planck2016-L}; \citealt{Ghosh16}) which have reported a decorrelation between polarized flux seen at 850\,$\mu$m and at 1.4\,mm (217\,GHz) in the diffuse ISM. While our current observations show imperfect correlations between polarized flux in three submillimeter bands with that seen at 850\,$\mu$m (see Figure \ref{fig:scatter}), at present we cannot distinguish a physical decorrelation from the scatter in our data due to statistical noise. We note, however, that if the observed millimeter decorrelation were due to variation in dust properties along the line of sight, our filtering method would likely make us insensitive to the effect. The high-pass filter was specifically chosen to isolate the translucent molecular cloud from lower spatial frequency components like 1/\textit{f} noise or diffuse Galactic emission. Future experiments, such as BLAST-TNG (\citealt{Galitzki14_TNG}; \citealt{Dober14}), will be necessary to describe variations in the polarization signal within a cloud, across many clouds, and over a broader range of ISM phases.

An additional source of tension between theory and observation has been seen by \citet{planck2014-XXI}. Those authors find that the observed ratio of submillimeter polarized flux to $V$-band polarized extinction degree ($P_S/p_V$) is larger by a factor of $\sim$2.5 than would be predicted by the DF09 models. \citet{Fanciullo17} found they were able to more closely match the empirical value of $P_S/p_V$ by modeling silicate grains with 20\% of their volumes replaced by vacuum inclusions. We have not explored the implications of models containing inhomogeneous dust grains in this paper, but if this type of modification effectively increases the temperature and the relative contribution to the emission from aligned silicate grains, the same solution might plausibly bring models into better agreement with the observed polarization spectrum. Similarly, \citet{Jones13} present a picture of dust evolution processes, invoking silicate grains with carbonaceous mantles and iron nanoparticle inclusions to match observations of unpolarized emission and extinction. Model grains like these may produce flatter submillimeter polarization spectra if they tend to reduce the temperature difference between aligned and unaligned populations.      

Modifications to models like DF09 models 1 and 3 that would produce a flatter polarization spectrum should focus on producing a population of aligned dust grains which have a temperature or spectral index more similar to the dust in the cloud as a whole. This could be accomplished, for example, by including a population of larger graphite grains, which could effectively reduce $\beta_g$ to be more similar to $\beta_s$. Perhaps more importantly for submillimeter wavelengths, the presence of larger grains overall would tend to eliminate the average temperature difference between silicate and graphite grains, making the average temperature of all grains more like the average temperature of aligned grains. We emphasize that these are suggestions that would tend to push models toward agreement with the observations we have discussed here. Rigorous models would need to fix their parameters based on constraints from other observations of extinction and emission and perform detailed calculations to evaluate predicted polarization spectra. 

Clearly there is a broad parameter space within which dust models may continue to be constrained. In this paper we have presented new observations that characterize submillimeter polarized dust emission in lower column-density environments than has been possible in the past. Certainly, more detailed modeling is necessary to separate the relative importance of the factors affecting grain alignment in marginally-shielded molecular clouds like the one we have observed. Our work presented here represents a new empirical constraint to be imposed on these future models.   

\acknowledgments{
The BLASTPol collaboration acknowledges support from NASA through grant numbers NAG5-12785, NAG5-13301, NNGO-6GI11G, NNX0-9AB98G, and the Illinois Space Grant Consortium, the Canadian Space Agency, the Leverhulme Trust through the Research Project Grant F/00 407/BN, Canada's Natural Sciences and Engineering Research Council, the Canada Foundation for Innovation, the Ontario Innovation Trust, and the US National Science Foundation Office of Polar Programs. FP thanks the European Commission under the Marie Sklodowska-Curie Actions within the H2020 program, Grant Agreement number: 658499 PolAME H2020-MSCA-IF-36  2014. JDS acknowledges the support from the European Research Council under the Horizon 2020 Framework Program via the ERC Consolidator Grant CSF-648505 and the European Community's Seventh Framework Programme FP7/2007-2013 Grant Agreements no. 306483 and no. 291294. LMF is a Jansky Fellow of the National Radio Astronomy Observatory (NRAO). NRAO is a facility of the National Science Foundation (NSF) operated under cooperative agreement by Associated Universities, Inc. ZYL is supported in part by NASA NNX14AB38G and NSF AST1313083. Based on observations obtained with Planck (\url{http://www.esa.int/Planck}), an ESA science mission with instruments and contributions directly funded by ESA Member States, NASA, and Canada. This research made use of data products from the Midcourse Space Experiment (MSX). Processing of the data was funded by the Ballistic Missile Defense Organization with additional support from NASA Office of Space Science. This research has also made use of the NASA/ IPAC Infrared Science Archive, which is operated by the Jet Propulsion Laboratory, California Institute of Technology, under contract with the National Aeronautics and Space Administration. This publication makes use of data products from the Wide-field Infrared Survey Explorer, which is a joint project of the University of California, Los Angeles, and the Jet Propulsion Laboratory/California Institute of Technology, funded by the National Aeronautics and Space Administration.
}

\appendix
\section{Analysis of systematic errors} \label{appendix_sys}

\subsection{Potential bias in fit method} \label{fitbias}
Here we examine the algorithm employed for performing the linear fits to the BLASTPol and \Planck FSN Stokes parameters, looking for evidence of potential biases. In particular, \texttt{MPFITEXY2} requires arrays of uncertainties on the $x$- and $y$-values for each point in the data set to be fit ($s_{x,i}, s_{y,i}$). We examine the effect on the resulting fit parameters when the estimates of errors supplied to \texttt{MPFITEXY2} are not equivalent to the ``true'' errors in the data ($\sigma_{x,i}, \sigma_{y,i}$).

As a test of this possible bias, we generate a simulated data set ($X_i$, $Y_i$) with
\begin{equation}\label{eq8}
X_i = x_i + f(\mu = 0; \sigma_x)
\end{equation}
\begin{equation}\label{eq9}
Y_i = x_i + f(\mu = 0; \sigma_y)
\end{equation}
where $x_i$ takes 1000 equally spaced values between -1 and 1, and  f($\mu$;$\sigma$) represents an independent sample from a normal distribution with mean $\mu$ and standard deviation $\sigma$. Intuitively, the simulated data sets represent perfectly correlated variables with a coefficient of unity, with varying amounts of Gaussian noise in each variable. We then fit for a slope, for convenience using the \texttt{MPFITEXY} algorithm \citep{Williams10}, which employs the same $\chi^2$ statistic minimization method as \texttt{MPFITEXY2}, but does not fit for the offset between two separate data sets. Any biases should be the same between the two methods, as the fitting algorithm is the same but with one fewer degree of freedom in \texttt{MPFITEXY}. With this arrangement, we can vary the true noise in the data as well as the noise reported to the fit algorithm and search for cases where the fit slope varies significantly from unity. 

In order to quantify the divergence of the fit slope from the injected slope of unity, we define an effective signal-to-noise ratio ($\Sigma$) in $x$ and $y$ as the range of the data in the absence of noise scaled by the width of the noise distribution: $\Sigma_x = 2/\sigma_x$ and $\Sigma_y = 2/\sigma_y$. We find that when the signal-to-noise ratio is marginal or better (e.g.\ $\Sigma \gtrsim$ 0.5) and the error bars used in the fit are accurate ($s_x = \sigma_x, s_y = \sigma_y$), the fit reliably recovers the true slope with a spread comparable to what would be expected from the fit's statistical errors alone. The same is true when $\Sigma$ is good but the errors in $x$ and $y$ have each been overestimated by the same factor ($s_x = k\sigma_x, s_y = k\sigma_y$). Even if the fit is performed assuming errors up to a factor of ten greater than the true $\sigma_x$ and $\sigma_y$ used to generate the simulated data, the resultant slope falls within the statistical error range. However, if the $\Sigma$ is marginal and the errors in one variable are overestimated more than the errors in the other ($s_x = k_x\sigma_x, s_y = k_y\sigma_y; k_x \neq k_y$), the fit slope will be significantly biased on an order larger than the statistical error bars. We find that for $\Sigma = 2$, errors in one variable can be overestimated by as much as 50\% relative to the other, and \texttt{MPFITEXY} will still return fit slopes and statistical errors that are consistent with the underlying noise-free relationship. This test gives us information on the cases for which \texttt{MPFITEXY} will provide unreliable fit parameters and statistical errors. If errors for $y$ are more overestimated than for $x$, fitted slopes will tend towards zero, while fitted slopes will tend to infinity in the alternate case.  

We know that the errors in our real data are overestimated (in either BLASTPol or \Planck or both) due to the fact that the reduced $\chi^2$ minimizes to a value less than unity when performing the linear fits (see Section \ref{stat_errors}). However, our tests show that as long as there is a significant amount of signal present (e.g., S/N $\gtrsim$ 2), the fits will not be biased by an amount larger than the reported fit errors. It can be seen from Figure \ref{fig:scatter} (left panel) that our S/N is of order 2, so the results reported in Table \ref{table:mainres} are not likely to be significantly affected by this bias. However, as spatial filtering can affect our signal strength, we will return to this issue in Section \ref{filt_errs} below.

\subsection{Other systematic errors} \label{other_errs}
\subsubsection{Systematic error due to spatial filtering} \label{filt_errs}
A significant source of systematic uncertainty in our analysis is the ambiguity in the proper values of the two Butterworth high-pass filter parameters, the cutoff $k_0$ and the order $n$ (see Equation (\ref{eq1})). Without an \textit{a priori} motivation towards a specific value for the two parameters, we instead take the approach of varying the parameters within a justifiable range. Then the systematic uncertainty is based on the variation of the polarization spectrum given that range of filtering parameters. Thus in this section we will also justify the values of the parameters used in the analysis presented in Section \ref{filt_method}. 

\begin{figure}
\includegraphics[width=1.1\columnwidth, center]{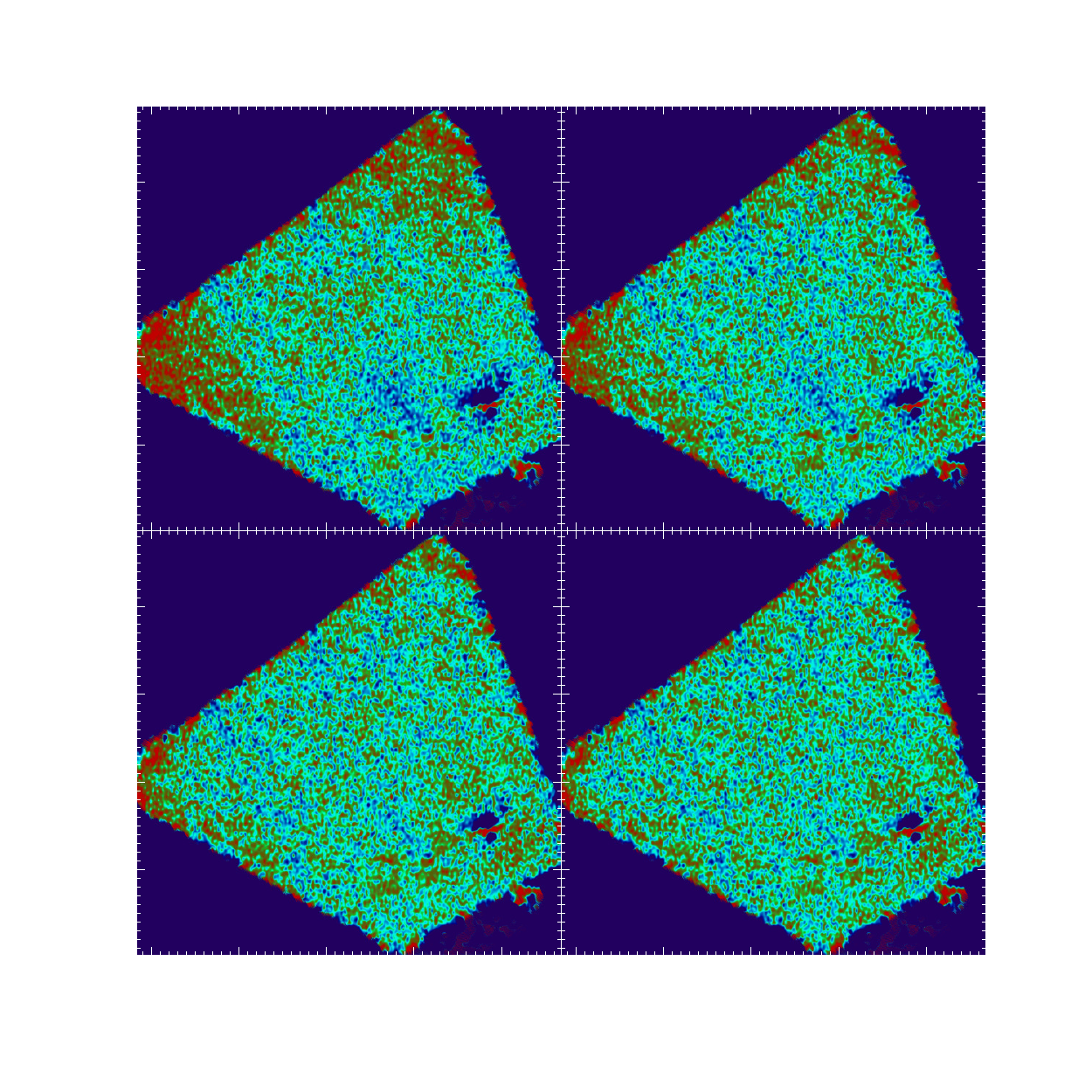}
\caption{Residuals of $Q_{250} - Q_{250}^{\FSN}$ maps after high-pass filtering with $k_0$ = 1, 2, 3, and 4. The color scale is the same as in Figure \ref{fig:bp_filt_resid}. \label{fig:cutoff_res}}
\end{figure}

\begin{figure}
\includegraphics[width=\columnwidth, center]{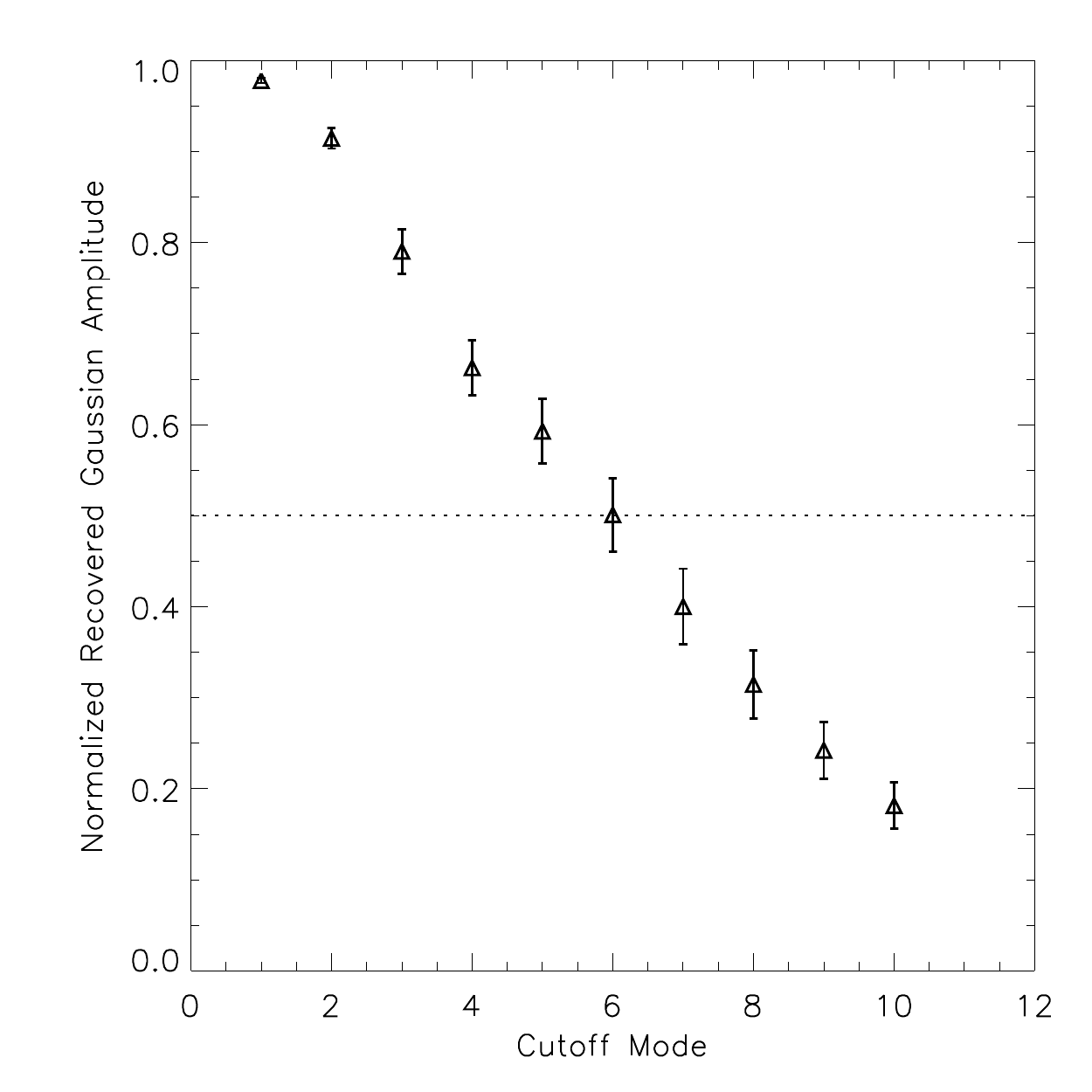}
\caption{Amplitude of a synthetic signal intended to emulate our target's $Q_{850}$, after filtering with a range of high-pass filter cutoffs $k_0$. Each triangular symbol represents the average amplitude under 11 independent noise realizations, and error bars represent 1-$\sigma$ variations. \label{fig:cutoff_max}}
\end{figure}

We begin by considering the cutoff mode, $k_0$. Our goal is to determine the smallest and largest values of the cutoff mode ($k_0^{min}$ and $k_0^{max}$, respectively) that could convincingly remove the contaminating 1/\textit{f} noise from the BLASTPol maps without removing signal on the scale of our translucent molecular cloud. In order to find $k_0^{min}$, we filter the BLASTPol and \Planck model maps using filters with n = 6 and successively larger $k_0$ values, starting with $k_0$ = 1 and increasing in integer steps. Maps of the (BLASTPol - Planck) residuals for $Q_{250}$ can be seen in Figure \ref{fig:cutoff_res}. For $k_0$ = 1 and 2, residual 1/\textit{f} noise can still be seen by eye in the corners of the map region. This artifact disappears for $k_0 > 2$, so we take $k_0^{min}$ = 3. The same trend holds for residuals in all three BLASTPol bands, for both $Q$ and $U$.
The constraint on $k_0^{max}$ comes from our observation above that fits to low-SNR data sets may have biased slopes (Section \ref{fitbias}). Thus it is important to have confidence that the filtering method employed is not removing an excessive amount of physical polarization signal on the scale of the target cloud. We approach this problem by creating simulated \Planck maps, as they generally have more noise than the corresponding BLASTPol map. Beginning with a map region of the same size as was used in the main analysis, we insert a 2D elliptical Gaussian with the same centroid position, widths, amplitude, and background level as seen in the \Planck FSN $Q$ maps. To this ``simulated signal" map we add random Gaussian noise drawn from a distribution with the same width as in low-signal regions of the Planck map. A single simulated map is then high-pass filtered with a series of Butterworth filters with $n$ = 6 and $k_0$ varying in integer steps from 1 to 10. We then fit a 2D Gaussian to the filtered maps, with the centroid and widths constrained to be equal to those of the injected signal but the Gaussian amplitude and background allowed to be free parameters. By comparing the amplitude of the injected signal with the amplitude recovered, we can describe the degree to which filtering at a given cutoff scale removes signal from a polarization structure. We repeat this process with 11 different realizations of the noise in the map, and show the result in Figure \ref{fig:cutoff_max}. We see that when $k_0$ = 6, under the majority of noise realizations, at least half of the signal amplitude was removed by the filtering. More signal is removed when $k_0$ is larger and the map is filtered on smaller scales, so we set $k_0^{max}$ = 5. 

\begin{figure}
\includegraphics[width=\columnwidth, center]{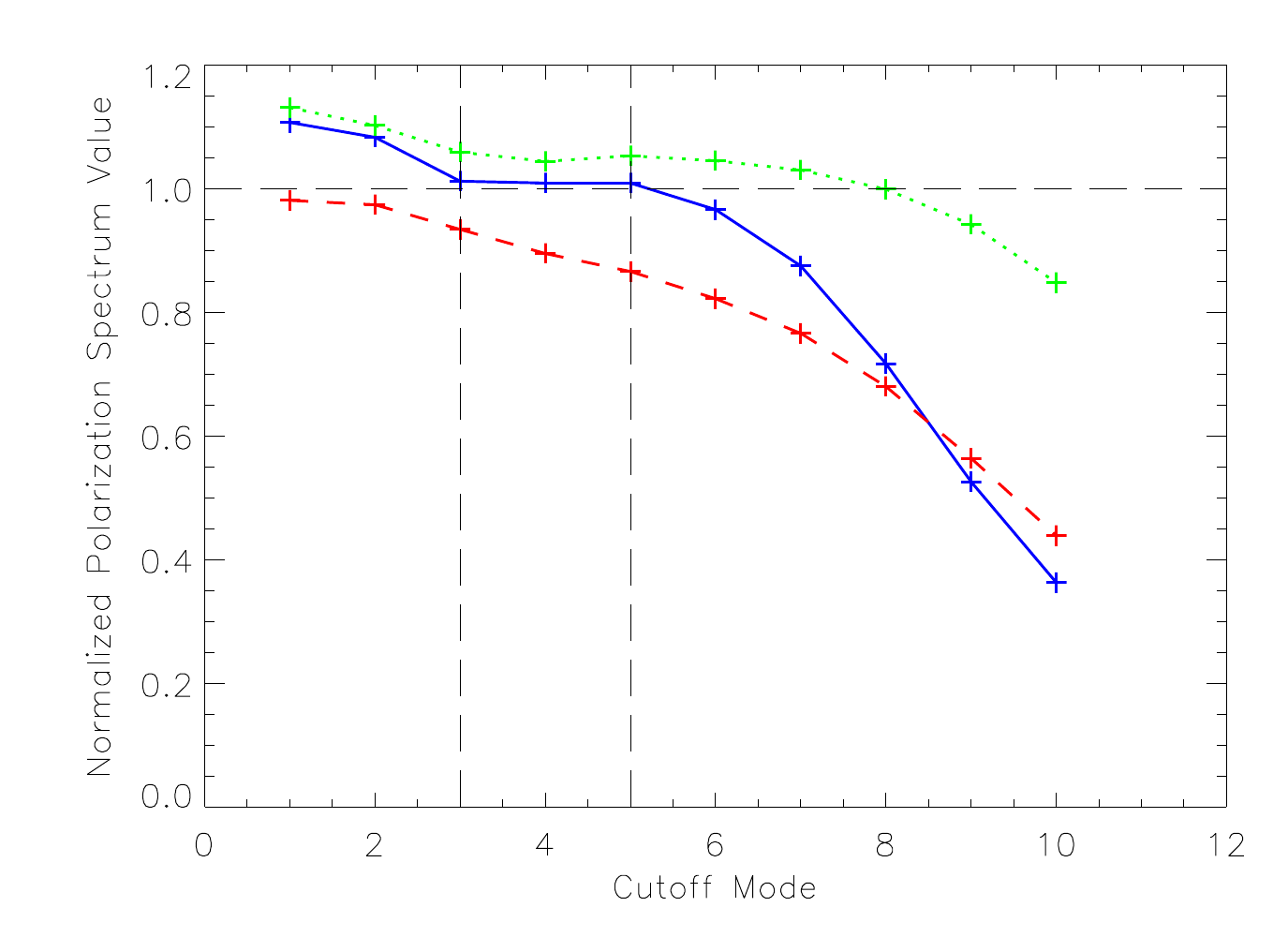}
\caption{Normalized polarization spectrum values $a_{\lambda}$ as a function of the high-pass filter cutoff $k_0$ for the BLASTPol 250\,$\mu$m (blue, solid), 350\,$\mu$m (green, dotted), and 500\,$\mu$m (red, dashed) bands. For reference, vertical lines show the range [$k_0^{min}, k_0^{max}$], and the horizontal line shows the value for a nominally flat polarization spectrum. \label{fig:polspec_cutoff}}
\end{figure}

Next, with the reasonable variations in the filter cutoff constrained to lie between $k_0$ = 3 and 5, we adopt $k_0$ = 4 for the main analysis (see Section \ref{filt_method}) and use the variation in the polarization spectrum after filtering within the range 3 $\leq k_0 \leq$ 5 to estimate the systematic uncertainty. Figure \ref{fig:polspec_cutoff} shows the variation in the normalized polarization spectrum for the three BLASTPol bands when the maps are filtered with a range of $k_0$ parameters. We see that for the 250\,$\mu$m and 350\,$\mu$m bands, there is very little variation within the range [$k_0^{min}$, $k_0^{max}$], which gives us confidence that our chosen cutoff scales are successfully removing only the 1/\textit{f} noise and not the cloud signal, as intended. The largest variation within this range of $k_0$ is seen in the 500\,$\mu$m band, which tends smoothly to lower values as $k_0$ increases. The values of the slopes fit after employing high-pass filters with $k_0$ = 3, 4, and 5 are shown in Table \ref{table:cutoff}. 

\begin{table*}[ht]
\begin{center}
\caption{Fitted Slopes $a_{\lambda}$ Versus Filter Cutoff}
\label{table:cutoff}
\begin{tabular}{cccc}
\hline
\hline
$\lambda$&$k_0 = 3$&$k_0 = 4$&$k_0 = 5$\\
\hline
250 $\mu$m&1.012&1.009&1.010\\
350 $\mu$m&1.059&1.044&1.053\\
500 $\mu$m&0.934&0.895&0.866\\
\hline
\end{tabular}
\end{center}
\end{table*}

\begin{table*}[ht]
\begin{center}
\caption{Fitted Slopes $a_{\lambda}$ Versus Order Parameter}
\label{table:order}
\begin{tabular}{cccc}
\hline
\hline
$\lambda$&n = 4&n = 6&n = 8\\
\hline
250 $\mu$m&0.999&1.009&1.008\\
350 $\mu$m&1.043&1.044&1.042\\
500 $\mu$m&0.882&0.895&0.898\\
\hline
\end{tabular}
\end{center}
\end{table*}

Finally, we investigate the polarization spectrum's sensitivity to the order parameter ($n$), which has been taken to be equal to 6 until this point. The order of the filter effectively controls the sharpness of the frequency cutoff, or equivalently, the amount of spatial power admitted from the stopband and removed from the passband. As we saw above, there is a fairly wide range of spatial frequencies over which the polarization spectrum is not very sensitive to the cutoff parameter. For this reason we expect our result to be relatively insensitive to variations in the choice of $n$, as this also amounts to varying the relative amount of power in modes 3 through 5 in the map. Our biggest concern is that the filter we use is sharp enough so that we effectively separate the low-frequency contamination from the cloud signal without significant leakage. As we established earlier that the information contained in mode 6 and above is important for describing the cloud polarization signal, we want to ensure that these modes are preserved by our choice of filter parameters. We note that when the cutoff mode $k_0$ = 4, if the order n = 6, then less than 1\% of the original power in mode 6 is removed by the filter. The polarization spectrum results when $k_0$ = 4 and n = 4, 6, and 8 are shown in Table \ref{table:order}. The differences are on the order of 0.01. As we anticipated, this variation is relatively small compared to that due to the choice of $k_0$.

\subsubsection{Systematic errors due to choice of region}
We check for systematic dependence of our result on the region around the cloud that is chosen to be sampled. We investigate the two most likely effects of the choice of the map region on the result: the proximity to the edge of the BLASTPol map and the size of the sampled region. It is possible that the high-pass filtering process introduced higher frequency ``ringing" artifacts into the filtered maps though we have made efforts to avoid the edges of the map where these effects would be present. To be sure this is the case, we repeat our analysis with the original Region A shifted $12.5^\prime$ (75 pixels) away from the edge of the BLASTPol map, keeping the polarized flux peak within the region (Region B; Figure \ref{fig:diff_reg}). The resulting polarization spectrum values are listed in Table \ref{table:regions}, and are consistent with our previous result to within the statistical error bars. Furthermore, to be sure that the inclusion of regions with little polarization signal is not diluting the signal and biasing the linear fits, we repeat our analysis with the selected region cropped to contain the polarized flux peak more tightly (Region C; Figure \ref{fig:diff_reg}). Again, the result of this test is listed in Table \ref{table:regions}, and it is consistent with our earlier result.

\begin{table*}[ht]
\begin{center}
\caption{Fitted Slopes $a_{\lambda}$ Versus Sampled Region}
\label{table:regions}
\begin{tabular}{cccc}
\hline
\hline
$\lambda$&Region A&Region B&Region C\\
\hline
250 $\mu$m&1.009&0.996&1.019\\
350 $\mu$m&1.044&0.996&1.043\\
500 $\mu$m&0.895&0.876&0.878\\
\hline
\end{tabular}
\end{center}
\end{table*}

\section{Two-Component Translucent Molecular Cloud Model Specification} \label{appendix_2compmdl}

\subsection{Extinction effect on temperatures}
To determine the radiation field that heats and aligns dust grains in each component, we use the IDL routine \texttt{CCM\_UNRED}\footnote[6]{ \url{https://idlastro.gsfc.nasa.gov/ftp/pro/astro/ccm\_unred.pro} }, which uses the extinction law of \citet{CCM89} with the update to the UV extinction by \citet{ODonnell94} to predict the relative reduction of intensity as a function of wavelength and extinction. Given a typical ratio of total to selective extinction ($R_V$) for the diffuse ISM of 3.1, we identify the wavelength at which the ISRF intensity will be reduced by a factor of 2 for each component of the cloud ($\lambda_{1/2}$). For the bulk and the surface, given the extinctions listed above, $\lambda_{1/2}$ is 0.4\,$\mu$m and 0.1\,$\mu$m, respectively. In the ``bulk-max'' case where the bulk is artificially set to have a characteristic extinction of 1.3\,mag, $\lambda_{1/2}$ is 0.8\,$\mu$m.   

To determine the dust temperatures as a function of grain size and material, we use the fact that dust temperatures $T \propto U_*^{1/6}$ \citep{Draine11} where $U_{*}$ is the starlight energy density\footnote[7]{We use the symbol $U_{*}$ here to eliminate possible confusion with the Stokes parameter $U$.} relative to that in the model by \citet{MMP83}, $u^* = 1.06 $ x $10^{-12}$ erg cm$^{-3}$. It is then necessary to calculate a value of $U_*$ in each of the two cloud components. \citet{MMP83} describe the spectrum of the ISRF as a combination of three dilute blackbodies at 3000 K, 4000 K, and 7500 K, with an additional UV component at 0.091 $\mu$m $< \lambda <$ 0.245 $\mu$m. Thus we specify $U_*$ by removing all contributions to the energy density at wavelengths shorter than $\lambda_{1/2}$ and scaling the result by the \citet{MMP83} starlight energy density. For the bulk, this results in a value of $U_*$ = 0.875, and accordingly the dust temperatures will be reduced by a factor of $(0.875)^{1/6} \approx 0.98$ relative to dust in the diffuse ISM. For the surface component, we note that $\lambda_{1/2}$ is approximately equal to the short-wavelength cutoff of the \texttt{CCM\_UNRED} domain (0.1 $\mu$m vs. 0.091 $\mu$m), so very little of the ISRF energy density will be removed. Thus we adopt $U_*$ = 1 in the surface, and the dust temperatures will be unchanged relative to the diffuse ISM. We apply these modifying factors to the temperature distributions of \citet{LD01}. In the bulk-max case, $U_*$ = 0.62, and the \citet{LD01} temperatures will be reduced by a factor of 0.92.  

\subsection{Extinction effect on grain alignment}
The minimum aligned grain size $a_{\mathrm{align}}$ is set to be  $\lambda_{1/2}$/2 = 0.2\,$\mu$m in the bulk, in accordance with the RATs prediction that alignment efficiency falls for $\lambda/a > 2$ \citep{LH07}. This is also the approach in the bulk-max case, where $a_{\mathrm{align}}$ = 0.4\,$\mu$m. In the surface, we leave $a_{\mathrm{align}}$ = $\lambda_{1/2}$ =  0.1 $\mu$m. This choice is made to avoid the inconsistency of using value of $a_{\mathrm{align}}$ smaller than was used in the diffuse ISM model of Section \ref{DFmdl}. This change to the size distribution of aligned grains will only introduce a small change in the calculated polarized flux, as the dust emission is dominated by the largest grains.

\bibliography{References,Planck_bib} 

\end{document}